%% file: sample-journal.tex
\documentclass[manuscript, review=false, screen]{acmart}
\usepackage{booktabs} 
\usepackage{subfig} 
\usepackage{color}
\usepackage[ruled]{algorithm2e} 
\graphicspath{{./figures/}}
\acmJournal{TOMM}
\acmVolume{9}
\acmNumber{4}
\acmArticle{39}
\acmYear{2018}
\acmMonth{3}

\setcopyright{acmcopyright}

\acmDOI{0000001.0000001}

\begin{document}
\title{Cloud Gaming With Foveated Graphics}

\author{Gazi Illahi}
\affiliation{%
  \institution{Aalto University}
  \city{Espoo}
  \postcode{02150}
  \country{Finland}}
\email{gazi.illahi@aalto.fi}

\author{Thomas Van Gemert}
\affiliation{%
  \institution{Aalto University}
  \city{Espoo}
  \postcode{02150}
  \country{Finland}
}
\email{thomas.vangemert@aalto.fi}

\author{Matti Siekkinen}
\affiliation{%
  \institution{Aalto University}
  \city{Espoo}
  \country{Finland}
}
\email{matti.siekkinen@aalto.fi}

\author{Enrico Masala}
\affiliation{%
  \institution{Politecnico di Torino}
  \city{Turin}
  \country{Italy}
}
\email{enrico.masala@polito.it}

\author{Antti Oulasvirta}
\affiliation{%
  \institution{Aalto University}
  \city{Espoo}
  \country{Finland}
}
\email{antti.oulasvirta@aalto.fi}

\author{Antti Yl{\"a} -J{\"a}{\"a}ski}
\affiliation{%
  \institution{Aalto University}
  \city{Espoo}
  \country{Finland}
}
\email{antti.yla-jaaski@aalto.fi}

\begin{abstract}
Cloud gaming enables playing high end games, originally designed for PC or game console setups, on low end devices, such as net-books and smartphones, by offloading graphics rendering to GPU powered cloud servers. 
However, transmitting the high end graphics requires a large amount of available network bandwidth, even though it is a compressed video stream. 
Foveated video encoding (FVE) reduces the bandwidth requirement 
by taking advantage of the non-uniform acuity of human visual system and by knowing where the user is looking. 
We have designed and implemented a system for cloud gaming with foveated graphics using a consumer grade real-time eye tracker and an open source cloud gaming platform. In this article, we describe the system and its evaluation through measurements with representative games from different genres to understand the effect of parameterization of the FVE scheme on bandwidth requirements and to understand its feasibility from the latency perspective. 
We also present results from a user study. The results suggest that it is possible to find a "sweet spot" for the encoding parameters so that the users hardly notice the presence of foveated encoding but at the same time the scheme yields most of the bandwidth savings achievable.
\end{abstract}

%
%

\ccsdesc[500]{Information systems~Multimedia streaming}
\ccsdesc[300]{Applied computing~Computer games}
\ccsdesc[300]{Computing methodologies~Image compression}
\ccsdesc[100]{Human-centered computing~User studies}

%
%

\keywords{Cloud Gaming, Foveated Video Encoding, Adaptive Bitrate Encoding
}

\maketitle

\input{introduction}
\input{background}

\input{prototype}
\input{evaluation}

\input{user_study}
\input{conclusion}

\bibliographystyle{ACM-Reference-Format}
\bibliography{biblio}
\end{document}

%% file: introduction.tex
\section{Introduction}
\label{sec:intro}
High-end gaming involves complex rendering of graphics, which is performed by dedicated GPU cards or chipsets on PC and game console setups. The graphics processing power of low-end PCs as well as net-books and smartphones is typically insufficient for such gaming. Cloud gaming makes high-end gaming possible on low-end devices by offloading graphics rendering to GPU powered cloud servers. The server intercepts rendered scenes, encodes them into a video, and streams the video to a thin client, which decodes and plays the received video. The client also intercepts user input and relays it to the cloud server where it is replayed locally. This kind of system allows remote game execution without having to modify the game or its underlying engine in any way. 

Cloud gaming infrastructure must satisfy the following constraints that are critical to the quality of experience (QoE) provided by the service: 1) short enough end-to-end latency between user input and corresponding change in video frame, and 2) large enough amount of available bandwidth to stream high quality video from the remote server. The first one stems from the fact that user perceived latency between control input and visible action on display reduces the quality of user experience in thin client computing generally~\cite{tolia06computer} as well as in cloud gaming specifically~\cite{Choy2012}. The second one arises from the need to stream sufficiently high quality video to the client device. The bitrate of a full HD video compressed using H.264 with typical settings would usually range from 5 to 10 Mbps, but it can go up to tens of Mbps depending on the encoder settings and framerate. Increasing the resolution to 4K would generally boost the bitrate up by at least a factor of three, which is, at the time of writing, reaching the limit for most of the Internet users\footnote{http://www.speedtest.net/global-index}. 

In this paper, we focus on the bandwidth challenge 
and propose to apply so called \textit{foveated video encoding} (FVE) to reduce the bandwidth requirement in cloud gaming. The method takes the non-uniform acuity of the human visual system (HVS) into account when encoding video: 
fovea is the region of the retina directly behind the eye lens. Visual acuity of the eye is the highest in the fovea and it drops sharply with angular distance from the fovea \cite{wandell}. Our approach is to encode game video rendered in the remote server with quality that spatially matches the acuity of HVS by tracking the gaze of the user and using the information during encoding. Encoding video in such a fashion can result in a significantly lower video bitrate compared to non-foveated encoding, hence reducing bandwidth requirement of cloud gaming.
Foveated video coding is an at least two decades old concept, which has drawn renewed interest in recent years because of affordable, good quality gaze trackers in the market and low latency networking technologies that have made new application scenarios possible. The difference of our approach compared to recent related work on cloud gaming (e.g., ~\cite{ahmadi14msys,mohammadi15icmew}) is that we apply real-time gaze tracking and foveated video coding to off-the-shelf games without need for game engine customization.  


The main contribution of this paper is to demonstrate that cloud gaming with real-time foveated graphics using commodity solutions available today can reduce the need for bandwidth by several tens of percent without the user perceiving any degradation in video quality. To this end, we have conducted a user study to examine the effect of foveated graphics of the user experience using a prototype system. While adjusting parameters in the foveated video encoding, we ask the users to rate video quality and their engagement. The study reveals that a "sweet spot" exists and may be found with careful parametrization of FVE, which yields large bandwidth savings with hardly any degradation in user experience. We also evaluate the prototype system with different parameter values and different games considering gaze data and video bitrate.


The paper is structured as follows. In Section \ref{sec:back} we discuss the background and related work of this work including cloud gaming, foveation and foveated video coding. In Section \ref{sec:system}, we describe the implementation of the prototype. In Section \ref{sec:eval},  we evaluate the system from bandwidth and latency perspectives and in Section  \ref{sec:user_study} we describe our user study and its results. Finally we conclude the work in  \ref{sec:conclusions}. 

%% file: background.tex
\section{Background and Related Work}
\label{sec:back}


\subsection{Cloud Gaming}
\label{sec:cloud_gaming}

\begin{figure}[ht]
 \begin{center}
   \includegraphics[width=0.6\linewidth]{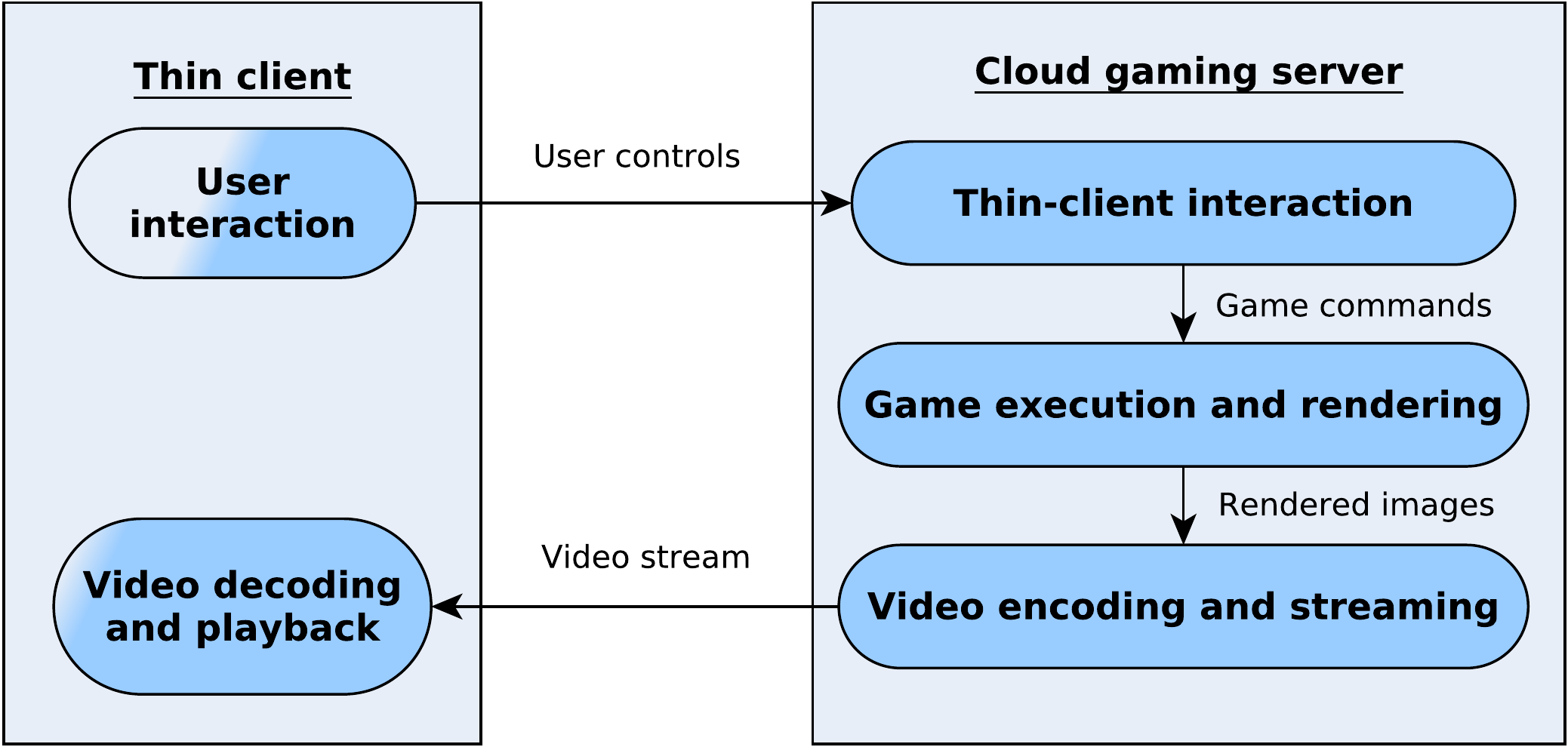}
   \caption{Cloud gaming architecture~\cite{kamarainen17mmsys}, \cite{Huang_GA}.}
   \label{fig:cloud_gaming}
 \end{center}
\end{figure}

Cloud gaming applies the cloud computing paradigm to gaming, dividing the implementation of a gaming system between a remote server and a local thin client, on which a player plays the game \cite{kamarainen17mmsys,shea2013cloud}. Typically, the cloud gaming server is deployed in a cloud or edge server as a virtual machine (VM) or a container. The cloud gaming server runs the game engine, captures game play video rendered by the engine, encodes it and transmits it to the cloud gaming client. The cloud gaming client receives the game play video, decodes and plays it out. The client also captures user input, such as key presses or mouse/joystick movements and transmits them to the server, which relays the user input to the game engine. The user inputs appear local to the game engine. This architecture, shown in Figure \ref{fig:cloud_gaming}, allows any off the shelf game to be played in a cloud gaming environment, allowing high quality gaming on even low end devices and mobile phones with low compute and energy resources. Other approaches of cloud gaming exist as well, for example, rendering load may be dynamically shared between the thin client and the server \cite{Cai_2013}. Another approach is to render at the server, but instead of a video stream, send video objects to the client \cite{mohammadi15icmew}. These approaches, however, require modifications to the game engine, which has to be done on a game by game basis and consequently cannot be used with off the shelf games. A survey on cloud gaming and cloud gaming architectures can be found in \cite{Cai_2014}.

For immersive QoE, the entire scheme has to be abstracted from the user, which in effect means that there should be no observable delay (end to end) for the user and the video displayed should be of sufficiently high quality which requires significant downstream bandwidth. This puts constraints on the design and deployment of a cloud gaming system as mentioned earlier.

\subsection{Foveation and Foveated Video Encoding (FVE)}
\label{sec:foveation}
\begin{figure}[ht]
 \begin{center}
   \includegraphics[width=0.6\linewidth]{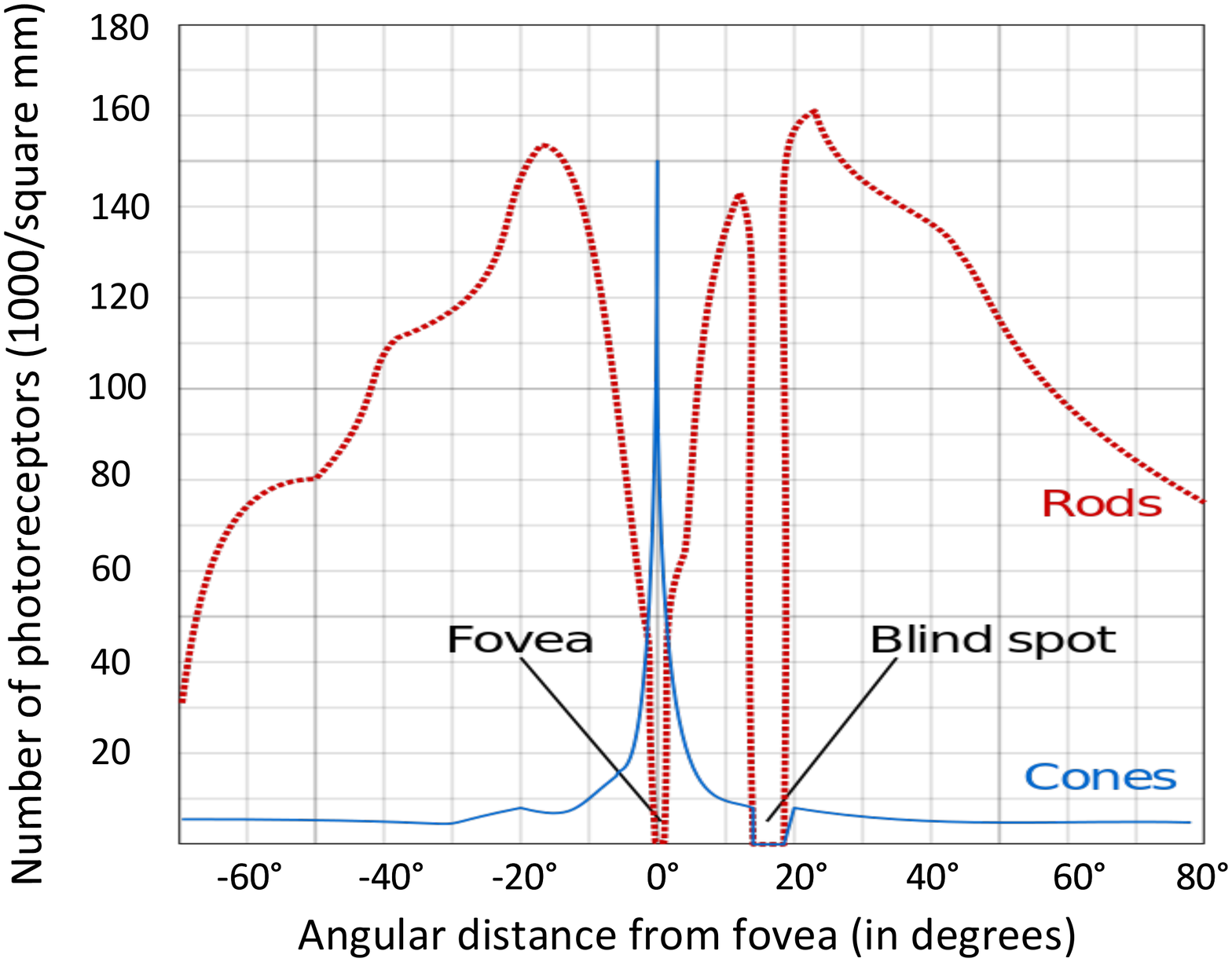}
   \caption{Density of photoreceptors in the human eye \cite{wandell}.}
   \label{fig:photo_receptor_density}
 \end{center}
\end{figure}

The human visual system (HVS) has a non-uniform sampling response to visual scenes, due to a phenomenon called foveation. Foveation is caused by the nature of distribution of photoreceptor cells in the human eye. There are two types of photoreceptors in our eye, the rods and the cones, responsible for vision under low illumination and high illumination levels, respectively~\cite{wandell}. The cones are involved in most activities like reading, gaming, etc. The density of photoreceptors in the retina is non-uniform, as shown in Figure \ref{fig:photo_receptor_density}. The cone density is highest at the fovea, which is the region of retina directly behind the lens, and drops off sharply as the distance from the fovea increases. Consequently, the sampling response of a visual scene and, hence, the perceived resolution corresponds to the density of cones, i.e., it is highest directly in front of the fovea and drops off sharply with angular distance from the fovea.
The acuity is highest within a range of 2\textsuperscript{o} of the human visual field \cite{ryoo16mmsys}. Beyond the 2\textsuperscript{o} angle the relative acuity falls off very sharply. Considering foveation, the practice of encoding video frame with  uniform spatial quality is wasteful. 

It should be noted that foveation is just one of the phenomenon involved in human vision. There are various pyscho-optical phenomena at play as well \cite{maintz2005digital}. Further, gaze fixation and consequently foveation is directly dependent on head and eye movements, of which, vestibulo-ocular reflexes, saccades and smooth pursuit are the most prominent. Vestibulo-ocular reflexes compensate for head movement when focusing on a point, saccades are fast eye movements in between fixations  and smooth pursuit occurs when the eye is tracking a moving subject \cite{Rand2006}. The interplay of foveation and eye movements is an actively researched topic \cite{shanidze2016accuracy} and is out of scope of this work. Here foveation is considered, which occurs during the so called fixations \cite{wandell}. Fixations occur when the eyes focus on a target, in between saccades. Further, there are various eye movements involved in maintaining visual fixation, for example, micro-saccades, ocular drifts and ocular micro-tremors \cite{rucci2016} which may affect eye tracking and hence eye tracking based FVE.

An effective real-time streaming scheme optimizes the trade-off between delivered video quality with respect to the available bandwidth by employing so called adaptive bitrate streaming. In adaptive bitrate streaming, bit rate of the streamed video, and hence its quality, is varied according to the available bandwidth. The change of quality is temporal, changing based on factors including network conditions. Foveated video streaming differs from traditional adaptive streaming in that the video quality may be changed spatially within a frame as well: encoding with  highest quality where the user gaze is fixated or predicted to fixate and with lower quality elsewhere. This spatial rate adaptation, used concurrently with temporal rate adaptation can yield significant improvements in streaming efficiency with respect to available bandwidth and delivered QoE.

Foveated video coding has been studied for quite some time, as noted earlier. A survey on the field was published by Wang et al.~\cite{wang2006foveated} about a dozen years ago. However, FVE has seen limited deployment, primarily due to the requirement of gaze information of each individual viewer for effective FVE. There are two approaches of determining the gaze location of a viewer, either by pre-analyzing the video for salient features where the user is likely to fixate their eyes or by tracking the gaze in real time. Analyzing video for salient features for foveated video coding has been an area of active research, for example, in \cite{bovik_2003,Itti_2004}.  The latter approach of using real time gaze location for foveated encoding has also drawn interest with availability of relatively economical non-invasive gaze tracking solutions. 
Most of the current approaches to FVE rely on tiling of the video frame. In \cite{DAcunto16mmsys}, the authors develop a prototype video streaming solution wherein the user can zoom in and navigate within a video. The prototype utilizes tiles together with the DASH-SRD extension \cite{niamut16mmsys}.The tiles which correspond to the zooming in location are delivered in higher resolution. Zare et al. describe a solution for VR applications using HEVC compliant tiles in ~\cite{zare16mm}. They partition 360\textdegree  video into HEVC compliant tiles with different resolutions and display high quality video tiles within the users view port. A similar solution is proposed by Qian et al. \cite{qian2016optimizing} for streaming panoramic video over wireless networks, streaming only the visible portion of the video. A foveated video streaming solution for video on demand is described by Ryoo et al. in \cite{ryoo16mmsys}, using real time web-cam based gaze tracking. Their approach is based on partitioning video into tiles and pre-coding videos into multiple resolution versions, streaming high resolution tiles at the gaze location and lower resolution tiles elsewhere.

\subsection{QoE}
\label{sec:Qoe_US}
QoE as a parameter, being subjective, is difficult to quantify. Although computational methods for assessing video quality have been around for a long time\cite{Akhtar2017}, with some supporting foveated video, the current system and experiment is focused more on the end-user experience. Objective measures may prove helpful in assessing the video quality, but we believe that the overall quality of experience is best reported by the users themselves \cite{Ijsselsteijn2007}. 
Many works have studied QoE in FVE and imaging, however, since a standard foveated encoding scheme does not exist, the parameters considered are different from the parameters we consider.
Lungaro et al.\cite{Lungaro2017} investigate a QoE of a similar FVE system in order to reduce the bandwidth requirements of high-resolution video. Their system defines a circular high quality region, an annular region around the foveal region with transitional quality and the background region with lowest quality.  The authors name the Round Trip Time (RTT, i.e. end-to-end latency) as the main constraint in such systems, and explore different combinations of encoding parameters (size and quality of the foveal, annular and background region) and network connection properties by employing a user study. The authors conclude that acceptable QoE may be achievable even with the current wireless networks with proper parametrization. Furthermore, they noticed that at some point increasing the size of the foveated area i.e. the area with high quality does not further increase QoE, and that instead increasing the background resolution (quality of the peripheral area) is needed.
Rai et al.\cite{Rai2017,Rai2016} explore the perceived video quality in foveated video systems, with a focus on artifacts in the peripheral area. In \cite{Rai2016},experimentals results indicate that non-flickering spatial artefacts in the peripheral region are less disruptive for the viewer than temporally flickering artefacts and also that the threshold of an artefact being diruptive is higher in visual periphery than in the foveal region. .The authors highlight the need for consideration of supra-threshold effects of distortions in the peripheral areas in order to maintain a high QoE. In \cite{Rai2017} the authors note that there may be a correlation between gaze disruptions and perceived video quality. By means of another user study they found a strong correlation (0.84) between gaze disruptions and DMOS (opinion scores).

In a cloud gaming system such as this a user's quality of experience will be largely determined by the perceived video quality and latency. How much so, especially with regard to latency, remains an area of active research. Some of our participants reported that latency issues were more disturbing than video issues. This is in line with results from other works like \cite{Claypool_latency}, but it should be noted that the extent of QoE deterioration  due to network delay in cloud and online gaming is highly dependent on the game genre and gameplay pace \cite{Jarschel_Qoe}. We refer the reader to \cite{Deber2015,Slivar2015,moller2013factors} for further reading on latency and QoE.

\subsection{Cloud Gaming and FVE}

Two works which focus on foveated graphics for cloud gaming are by Ahmadi et al.~\cite{ahmadi14msys} and Mohammadi et al.~\cite{mohammadi15icmew}. Ahmadi et al. ~\cite{ahmadi14msys}, develop a game attention model based on both saliency of the game play video and and a game priority model which considers objects in the video frame based on likelihood of game player's attention. Mohammadi et al. describe a solution to reduce downstream bandwidth required for cloud gaming. The authors use foveated encoding of graphics objects based on live gaze data. The key difference with our solution is that their solution uses MPEG 4 BiFS framework instead of streaming video directly and requires hooks into the game engine to get the game object data. These approaches require either prior knowledge of the game or changes to game engine or both, hence they cannot be used with off the shelf games. 

In our previous work \cite{Illahi2017}, we develop a foveated video streaming solution for cloud gaming, wherein we use a gaze tracker to track the gaze of a player in real time and send the gaze information to the cloud gaming server. At the cloud gaming server, we incorporate the gaze information in the gameplay video encoding scheme for foveated video encoding. The approach works with any off-the-shelf game and requires no modifications to the game engine. In \cite{Illahi2017}, we make optimistic but cautious conclusions about the feasibility of FVE in cloud gaming and the potential of bandwidth savings without affecting QoE. In this work, we extend our previous work by validating our observations on feasibility of FVE and potential bandwidth savings with a user study.

%% file: prototype.tex
\section{Cloud Gaming With Foveated Graphics}
\label{sec:system}
To implement FVE for cloud gaming, we use GamingAnywhere \cite{Huang_GA} as our cloud gaming platform. It is an open source, portable and extensible software comprising cloud gaming server and client implementations, optimized for performance. Further, to track gaze locations of a player we use an eye tracker on the client side. 
\begin{figure}[ht]
 \begin{center}
   \includegraphics[width=0.6\linewidth]{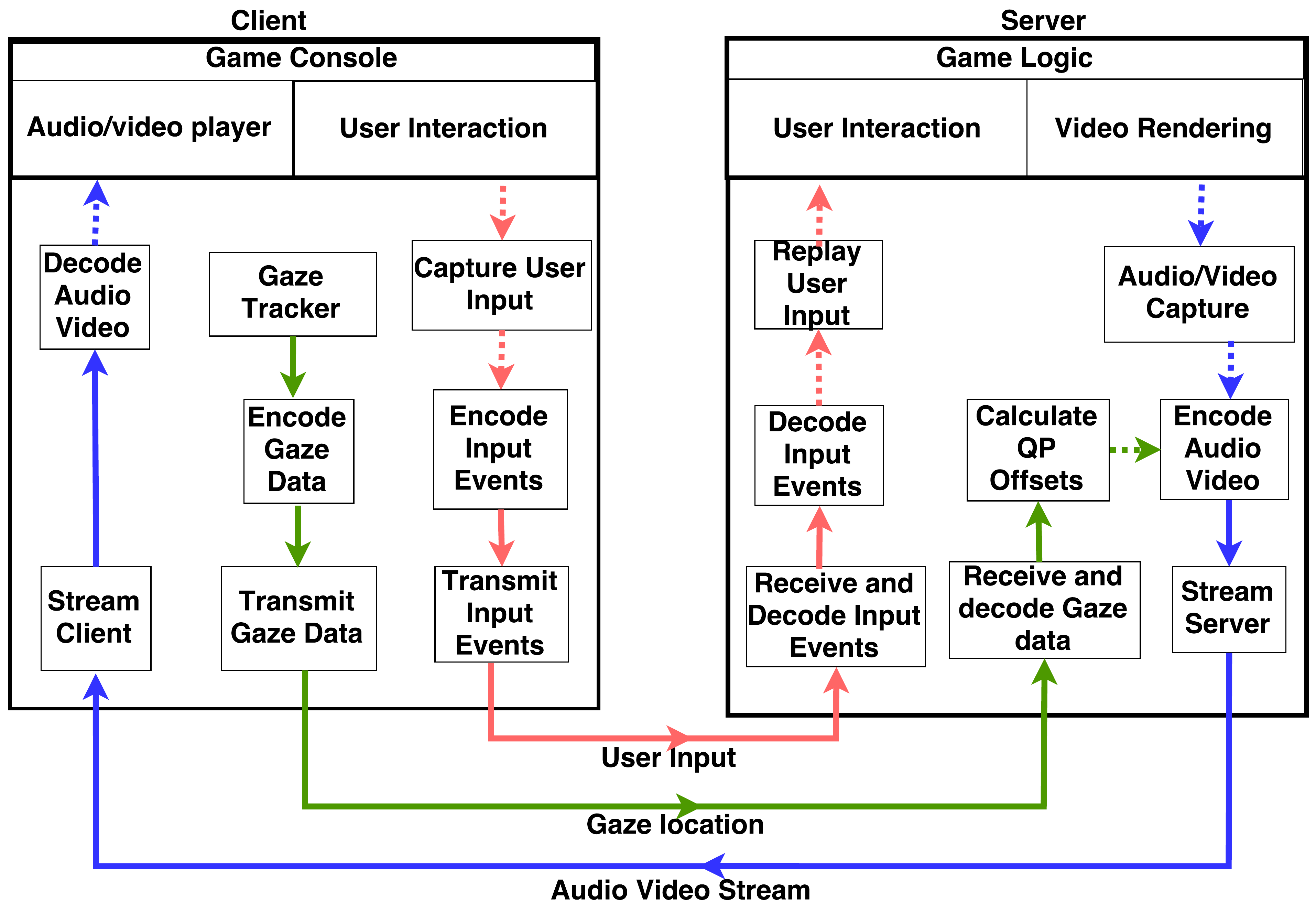}
   \caption{Architectural overview of the prototype}
   \label{fig:prototype}
 \end{center}
\end{figure}
The GamingAnywhere server captures game play video, encodes it and streams it to the GamingAnywhere client. It also receives user input from the client and replays them to the game engine. In our prototype, it is modified to accept real time gaze location data from the client. The gaze location is used by the server to encode video in a foveated fashion in real time. 
The GamingAnywhere client receives gameplay video from the server, decodes it and plays it out. It also captures user input actions, like key presses or mouse movements and forwards them to the GamingAnywhere server. In our prototype, a gaze tracker is also installed on the client machine, which tracks the user's gaze. The gaze location data is sent to the server as soon as it is available at the client, for seamlessness of encoding.\\
We use a Tobii 4C gaze tracker \footnote{ \url{ https://tobiigaming.com/eye-tracker-4c/}} to track the gaze and configure the GamingAnywhere server to use the {\em x264} encoder  with a preset {\em crf} in adaptive quantization mode. Adaptive quantization allows us to change the quantization parameter on a macroblock basis for each frame. The exact method of calculating the QPs is discussed in detail in \ref{sec:encoder_tracker_interface}. An architectural overview of the prototype is illustrated in Figure \ref{fig:prototype}.

\subsection{Gaze Tracker}
\label{sec:gaze_tracker}
Tobii 4C Eye Tracker  used at the client of the prototype is an economical eye tracking device directed towards gaming and human computer interaction applications. It has an on-board application specific integrated circuit which can track each eye, providing eye location, gaze location and other related data, invariant to head movements. We use the eye tracker in a "light filtering" mode where the gaze data is adaptively filtered considering both age and velocity of the reported gaze points, hence removing noise \cite{TobbiDevGuide}. Noise in gaze data may include, for example, micro-saccades which happen when the eyes are trying to focus on a target. We make a design decision to use lightly filtered gaze data instead of fixation data which is also extractable from the eye tracker, because the algorithm used by Tobii for fixation calculation is not publicly available. The lightly filtered gaze data as received from the eye tracker is encoded and sent over a TCP link to the server. The TCP link is parallel to, rather than coupled with, the user input channel and the video stream channel to prevent the other data flows from hindering the gaze data flow. The encoding is minimal and only adds timestamps to the gaze data. The server, as mentioned above, accepts the gaze data, decodes and sanity checks it to use only the latest gaze updates. The gaze coordinates are then used to develop a quality profile for the current frame in the encoding pipeline. 

\subsection{FVE with Real Time Gaze Tracking}
\label{sec:encoder_tracker_interface}
\begin{figure}[ht]
\centering
\subfloat[Visual acuity of human eye]{\label{fig:visual}\includegraphics[width=0.44\linewidth]{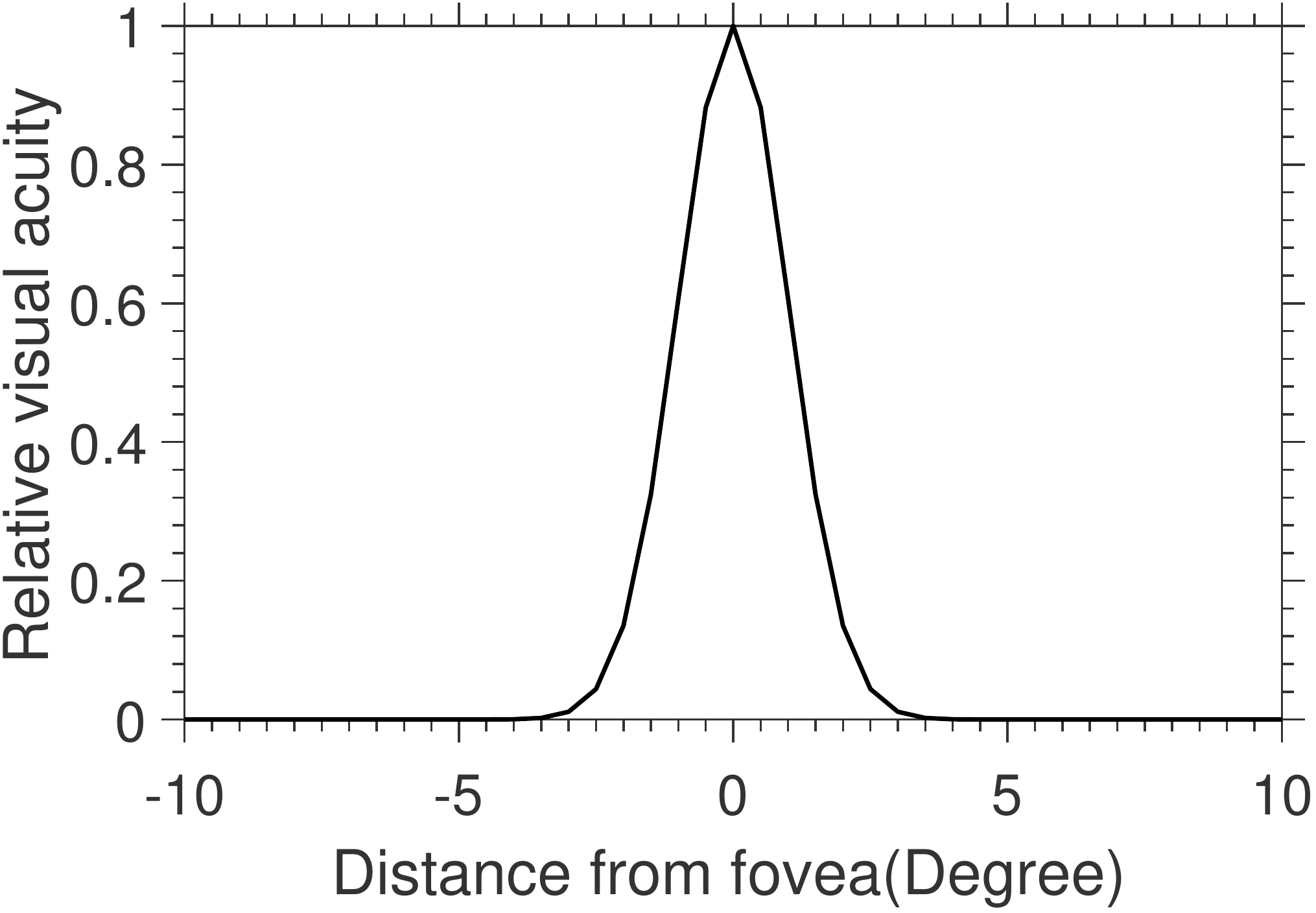}}\hfill
\subfloat[$QO$ at different values of $W$]{\label{fig:foveation}\includegraphics[clip, trim=0cm 0cm 0cm 0.3cm,width=0.45\linewidth]{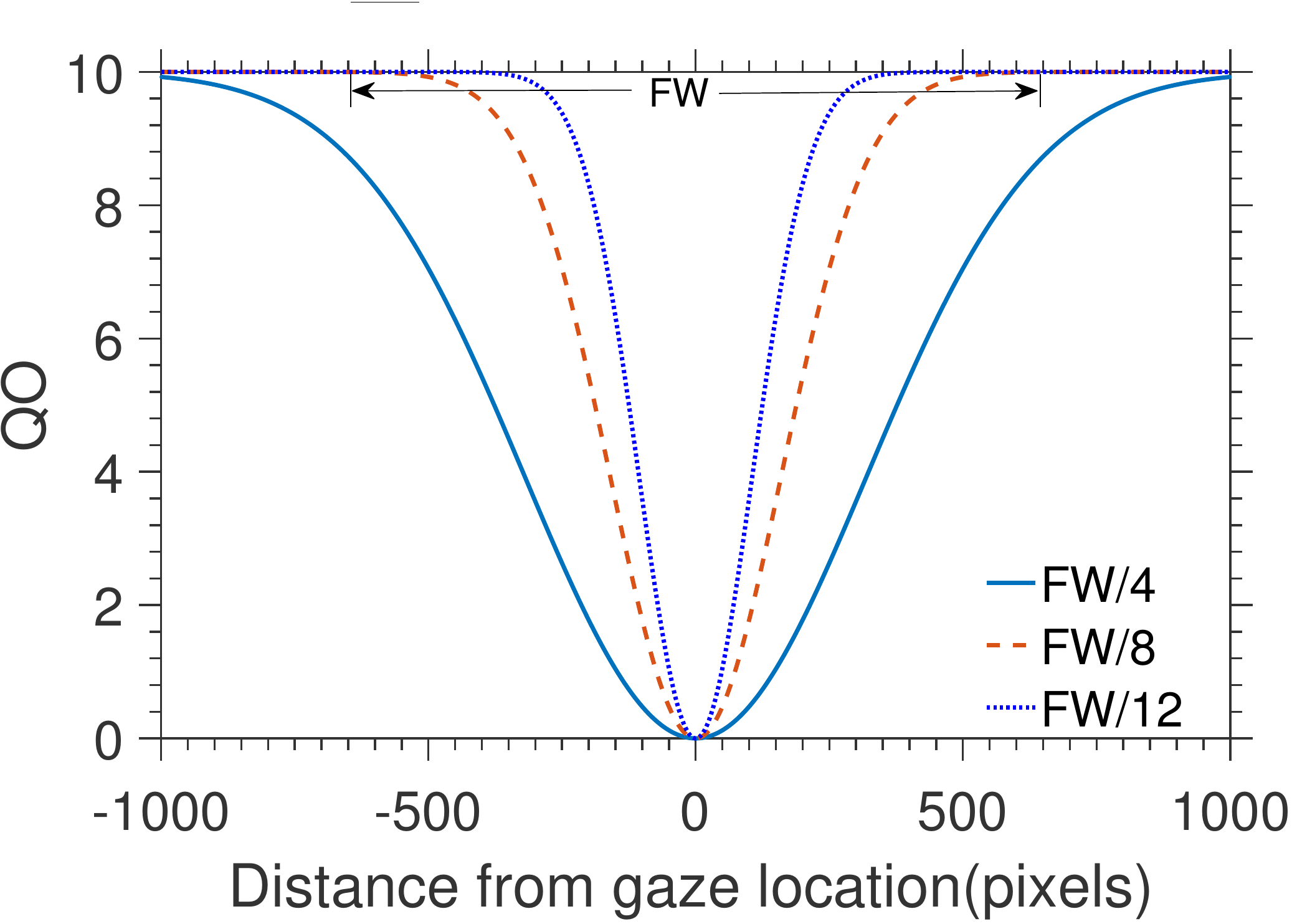}}
\caption{Foveation and $QO$ calculation. $FW$ is the width of the output frame in pixels}
\end{figure}

Foveation occurs because of the non-uniform density of cone cells in the human eye, as discussed earlier. The relative visual acuity of the human eye is illustrated in Figure \ref{fig:visual}. We simplistically model the relative visual acuity of the HVS as a two dimensional Gaussian function centered around the fovea. More complex models of foveation and relative visual acuity of the HVS for foveated encoding have also been developed, for example, in \cite{Chen_2010}. We make a design decision to use a simpler model to minimize the modifications needed in the video processing pipeline of the encoding scheme and to minimize latency overhead added by foveated encoding.

Video encoding has multiple steps of compression, which may be lossless or lossy. The primary lossy compression stage in modern encoders is the quantization step. This step of encoding enables actuation of the trade-off between quality and video bitrate. The level of quantization determines the quality of reconstructed video: the higher the quantization, the lower the quality and the resulting bitrate. Encoders may use different rate control strategies to achieve an optimal trade-off between quality and video bitrate by controlling quantization and other encoding parameters. Some strategies are, for example, constant quantization, attempting constant delivered quality or enforcing a constant bit rate. 
 
\subsubsection{Encoder} In this work the {\em x264} implementation \cite{x264} of Advanced Video Coding (AVC) standard \cite{h264:standard} is used. In {\em x264} the quantization levels are controlled by a parameter called Quantization Parameter (QP). QP is calculated for each macroblock. Depending upon the rate control strategies used, the QPs may be determined automatically by the encoder or set by the user. One of the rate control strategies available in {\em x264} is the so called constant rate factor (crf) mode, in which the encoder attempts to maintain a constant perceived quality temporally, determining QP values accordingly. In the {\em crf} mode, the encoder takes advantage of the fact that the human eye perceives still and moving objects differently and compresses the video according to the motion in the frame. For single pass encoding, this mode is considered the most efficient. To implement foveated encoding, we use the {\em x264} in the  {\em crf} mode, but we add an offset to the QPs calculated by the {\em crf}  algorithm. The QP offsets are calculated such that the offset is lowest at the gaze location and increases away from the gaze location. This scheme keeps the quality highest at the gaze location and lowers it away from the gaze.

\subsubsection{QP calculation}
In the prototype, the server is modified to accept gaze data sent by the client over a TCP connection parallel to the gameplay video and user input channels.
 At the server, the module responsible for inputting game play frames to the encoder calculates QP offsets for each macro block of the video frame. GamingAnywhere server and client negotiate gameplay video resolution when they connect initially, so the video processing modules know the number of macroblocks to expect. To calculate the QP offsets of each macro block, the gaze location is translated to a macroblock based coordinate system. The macroblock corresponding to the current gaze location is assigned the lowest QP offset, while macroblocks away from this macroblock QP offsets which increase progressively with distance from the gaze macroblock. Since we model the HVS acuity as a two dimensional Gaussian function, we calculate the offsets using a two dimensional Gaussian function. For the current video frame to be encoded, the QP offset, $QO(i,j)$ for a macroblock at $i , $j, where $i$ and $j$ are indices of the matrix of macroblocks comprising the frame, is calculated as:  
\begin{equation}
QO(i,j) = QO_{max}\left(1-\exp-(\frac{(i-x)^2+(j-y)^2}{2(W)^2})\right)
\label{eq:offsets}
\end{equation}
In \eqref{eq:offsets}, $QO_{max}$ is the maximum offset which is configurable by the server administrator (or user), $x$ and $y$ are the indices of the macroblock corresponding to the gaze location, and $W$ is a measure of the size of the foveal region. We define foveal region as the region on the screen of the client machine which corresponds to the gaze location and where the game video quality should be high.  $QO_{max}$ and $W$ are user configurable and allow us to investigate the relationship between quality offsets size of foveal region and video bitrate and QoE.

Figure \ref{fig:foveation} shows the relationship between $QO$ and distance from gaze  at various values of $W$ being varied in terms of the output frame width $FW$. In a video frame, the area perceived with the highest visual acuity depends on the viewing distance, larger viewing distance translates to a larger area sampled with high resolution by the HVS. In our prototype evaluations, we vary $W$ in terms of the frame width $FW$ because humans naturally tend to view smaller screens from a closer distance and larger screens from further distances. It should be noted that in this prototype $W$ represents the distance from the gaze center in frame to a circle where the QP offset is about 40\% of $QO_{max}$, following our simple model of HVS. We believe varying quality according to a Gaussian curve follows the HVS acuity better than step functions of quality variations used in other approaches of foveated encoding for example in \cite{ryoo16mmsys} and \cite{bokani2014empirical}.

%% file: evaluation.tex
\section{System Evaluation}
\label{sec:eval}
In this section, we investigate the effectiveness of foveated video streaming to reduce bandwidth requirements. We consider three games of different genres for analysis of their video bitrate with different parameterization of foveated encoding. Further, we briefly analyze player gaze patterns with four games to roughly estimate latency between a detected eye movement to change in the received video.
\subsection{Measurement Setup}
The measurement setup comprises of our cloud gaming system prototype as described in \ref{sec:system}, wherein the client and the server are connected over a campus GbE network, to minimize network related biases in the measurements. An Ubuntu work station serves as the server and a Windows laptop serves as the client. Three games of different genres with different game plays are considered for video bitrate analysis: AssaultCube, Trine2 and Little Racers STREET (henceforth abbreviated as Little Racers). %
AssaultCube is an action game of the First Person Shooter (FPS) genre, wherein the player controls a weapon from a selection of weapons from a first person point of view. Being an FPS game, it has a fast paced gameplay. Trine2 is a side scrolling puzzle and adventure game, wherein the player assumes one of a selection of in-game avatars and explores the gameplay map, solving challenges along the way. Little Racers is a so called top down car racing game wherein the player races a car from bird's-eye perspective on different race track maps available in the game.

We conduct a set of measurements for each game by capturing all traffic flowing between the cloud gaming server and the client using tcpdump \footnote{http://www.tcpdump.org/}. Raw tcpdump data is then analyzed with Wireshark \footnote{https://www.wireshark.org/} to extract throughput per second. In a set of measurements for a game, a player familiar with the gameplay controls, plays the game for a fixed period session for each set of parameters, making an effort to replicate gameplay over the sessions. To encode the gameplay video, the cloud gaming server is configured to use the {\em x264} encoder with the following parameters: \\
\texttt{--profile main --preset ultrafast --tune zerolatency
--crf 28 --aq-mode 1 --ref 1 --me\_method dia --me\_range 16 --keyint 48
--intra-refresh  --threads 4} \\
The encoding parameters are partly based on the suggestions provided by developers of GamingAnywhere in ~\cite{Huang2013ga}: 
\subsection{Foveation and Video Bitrate}
\label{sec:Fov_throughput}
To evaluate the effect of foveation on the video bitrate and consequent savings in the downstream bandwidth requirements, we perform a series of measurements with the above described setup. Over a set of measurements for each game, we vary the maximum offset $QO_{max}$ which controls the quality degradation keeping the $W$ parameter that controls the foveal region constant followed by varying $W$ while keeping $QO_{max}$ constant. As mentioned earlier, we define values of $W$ relative to the frame width ($FW$) of output video. Considering the fact that viewing distance generally increases with display size, this approach provides screen size agnostic visual quality adaptation.
\begin{figure}[ht]
\begin{center}
\subfloat[AssaultCube]{\includegraphics[width=0.5\linewidth]{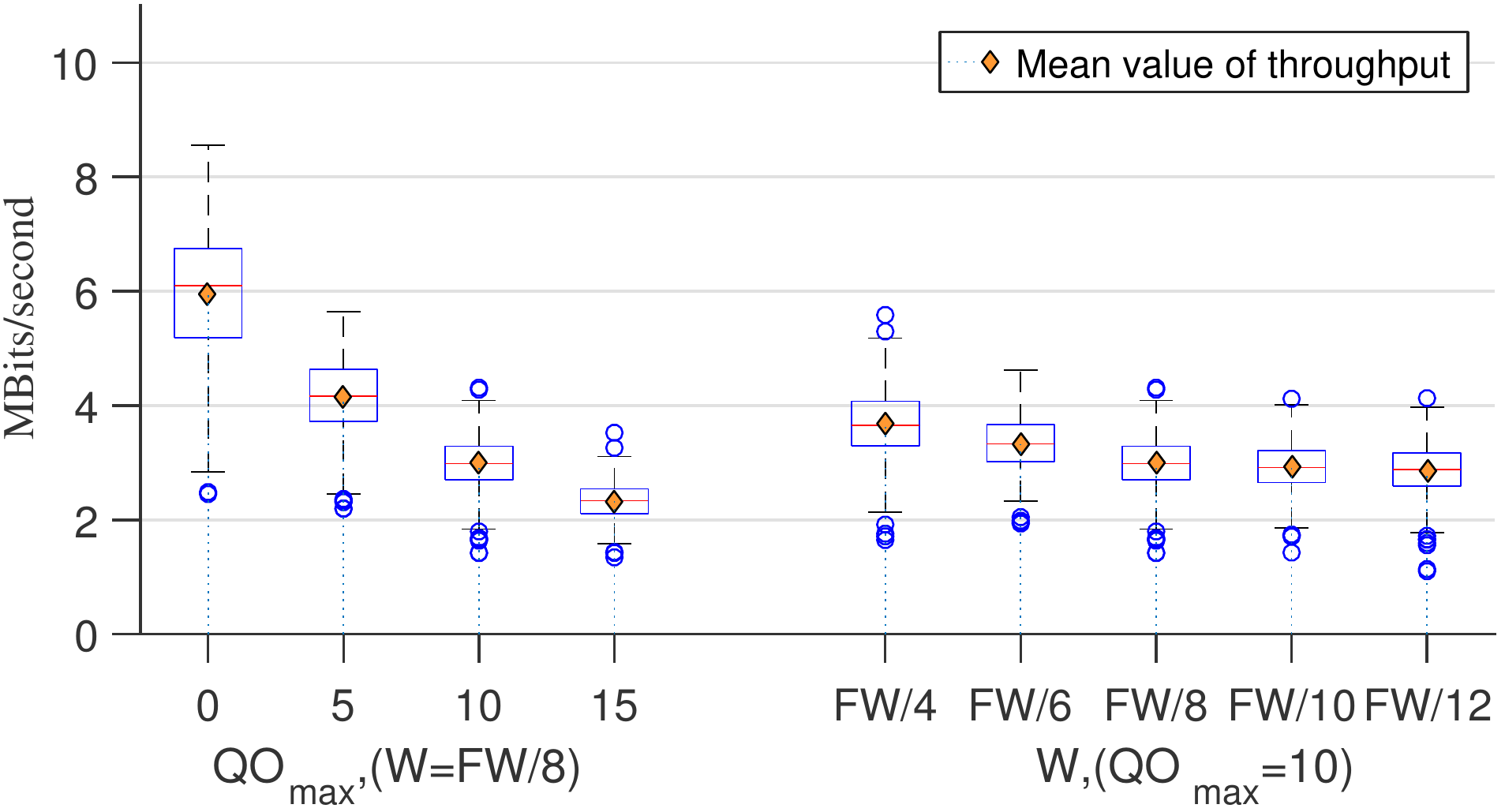}}\hfill
\subfloat[Trine 2]{\includegraphics[width=0.5\linewidth]{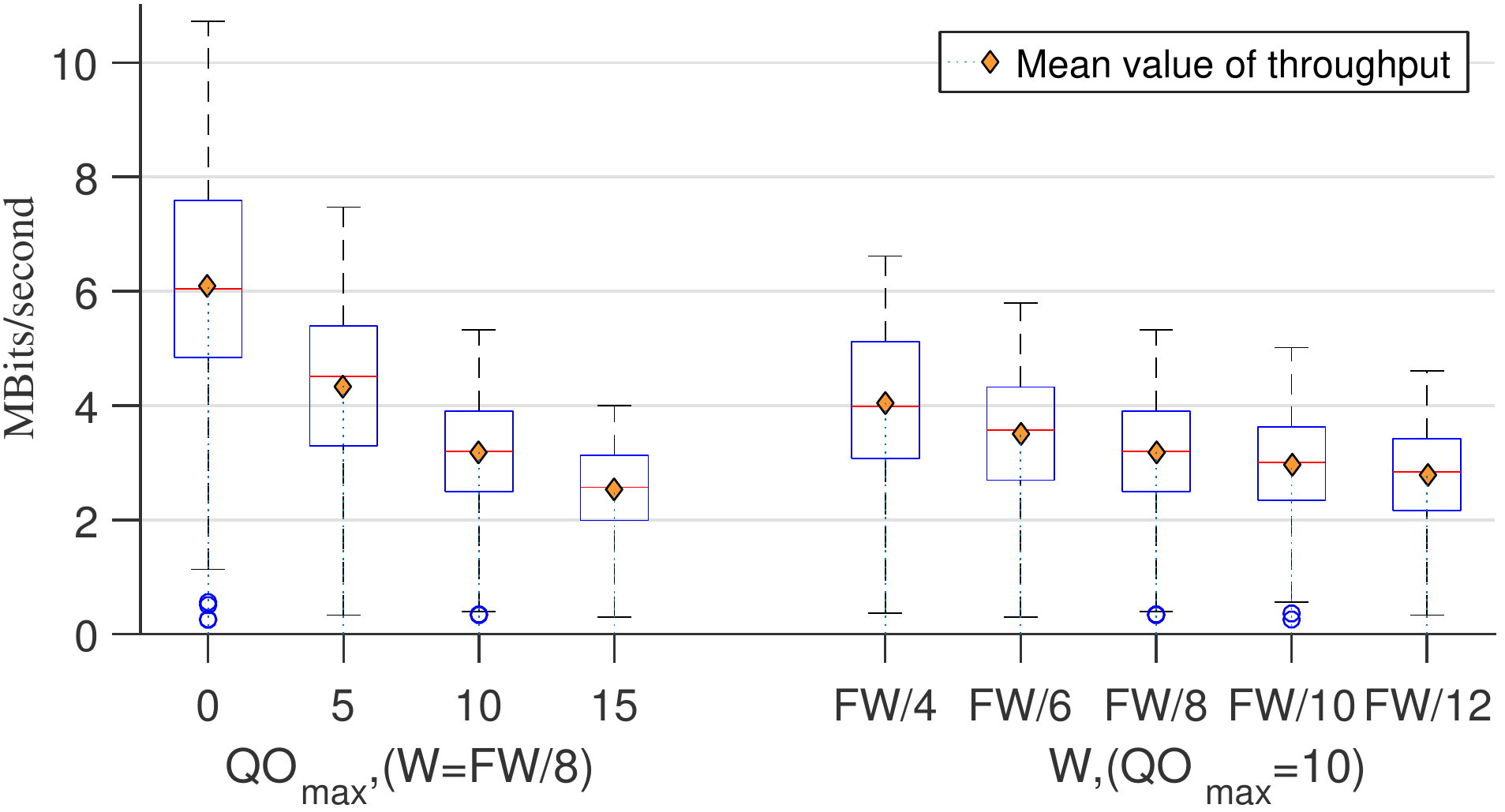}}\hfill
\subfloat[Little Racers STREET]{\includegraphics[width=0.5\linewidth]{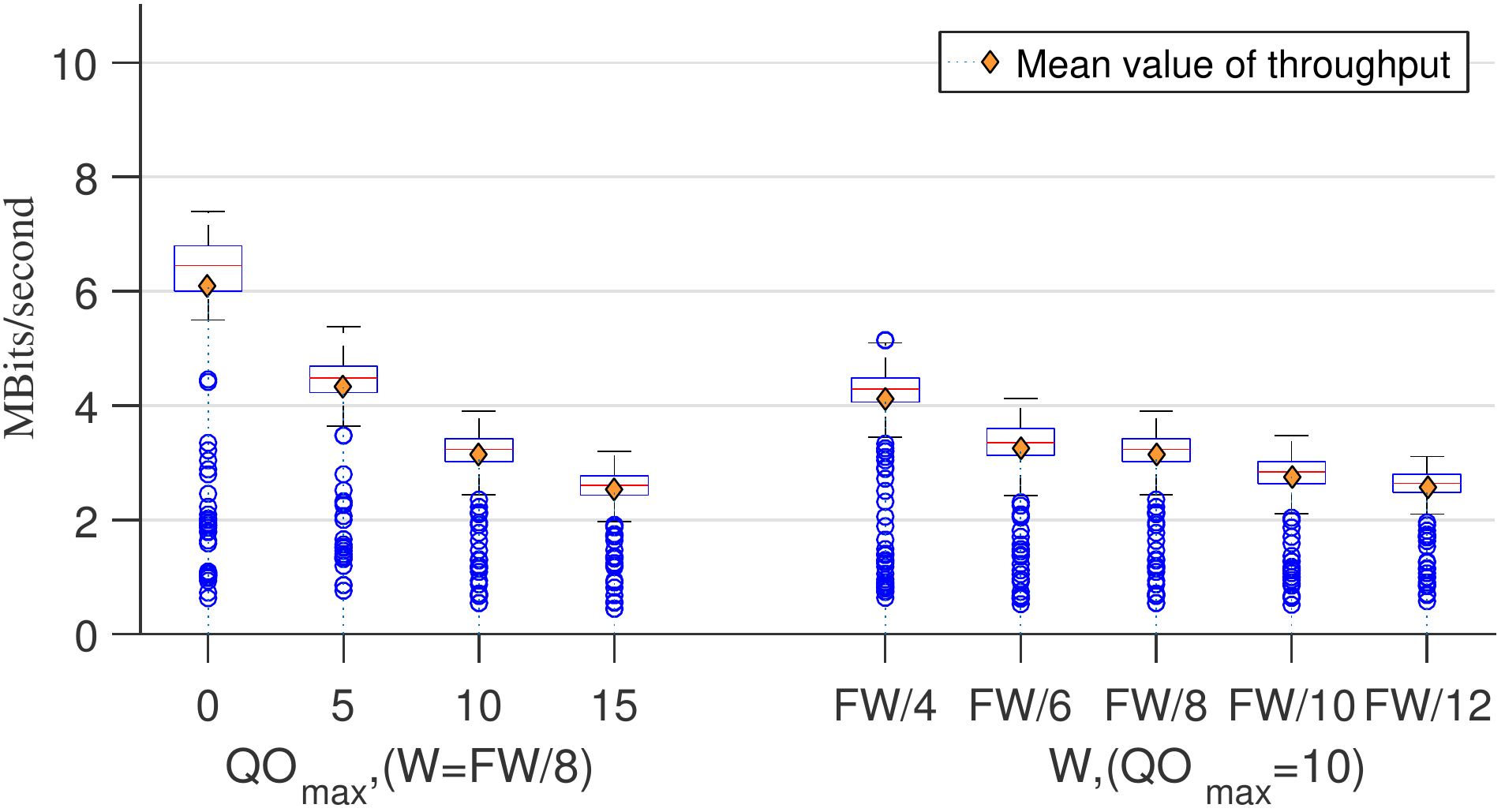}}
\caption{Measured video bitrates with different games and parametrization of foveated streaming. $FW$ is the width of the display in pixels. The box comprises the inter-quartile range, the red line in the middle of the box is the median, and the diamond denotes the mean.}
\label{fig:bitrates}
\end{center}
\vspace{-5mm}
\end{figure}
The results for the three games considered in throughput analysis are shown in Figure \ref{fig:bitrates}. Increasing the maximum QP offset $QO_{max}$, while keeping $W$ constant results in a significant change in the output bitrate. However, decreasing $W$ beyond 1/8th of the output frame size has no marked effect on output video bitrate, the reason being that beyond 1/8 of screen size  on the cloud gaming client of our setup, the number of macroblocks encoded at high quality (with low QP offsets) is a very small fraction of the total number of macroblocks on the screen. In the measurement setup the output frame width is 1366x768 pixels which corresponds to about 4000 macroblocks of 16x16 pixels while as with $W$= $FW/8$, the number of macroblocks with $QO$ offset of 40\% or less of $QO_{max}$ is about 100. At higher resolutions of output game play video, we expect the pattern of bandwidth savings to be similar to the discussion above, however, foveal regions of smaller size (below $W=FW/8$) should have a more pronounced effect as the number of macroblocks affected increases.

Comparing the results from the three games, it can be noted that there is little difference between the average or median bitrates in contrast to the variance which is significant. This is due to the nature of gameplay in each game. The least variance is in Little Racers which has a birds eye view perspective wherein even when the player controlled car is changing position constantly, the overall map and hence the frame graphics change infrequently. Trine2 which has complex graphics and where frame changes almost constantly with player actions, exhibits the most variance. However, changing the foveation parameters affects all the games in a similar fashion and it is safe to conclude that the bandwidth savings due to foveated encoding are game agnostic.

\subsection{Gaze Patterns and Latency}
\label{sec:gaze_latency}
To investigate the latency feasibility of foveated streaming for cloud gaming we study gaze data from four games. Three of the games are same as considered in section \ref{sec:Fov_throughput}. An additional game, Formula Fusion, is also considered. Formula Fusion is a "futuristic" racing game with a fast game play. The player point of view may be configured to be behind the vehicle or from inside the vehicle. We configure it in the behind the vehicle mode. For the analysis, the Tobii 4C eye tracker is configured to capture gaze data for each game while a player plays the game on a Windows computer, each game being played for 15 minutes. Sample screen captures of the games are presented in Figure \ref{fig:screencaps}. From Figure \ref{fig:screencaps}, we can observe where the user is likely to fixate their gaze and where they might glance occasionally. 

\subsubsection{Gaze Patterns}
\begin{figure}[ht]
\begin{center}
\subfloat[AssaultCube]{\includegraphics[width=0.5\linewidth]{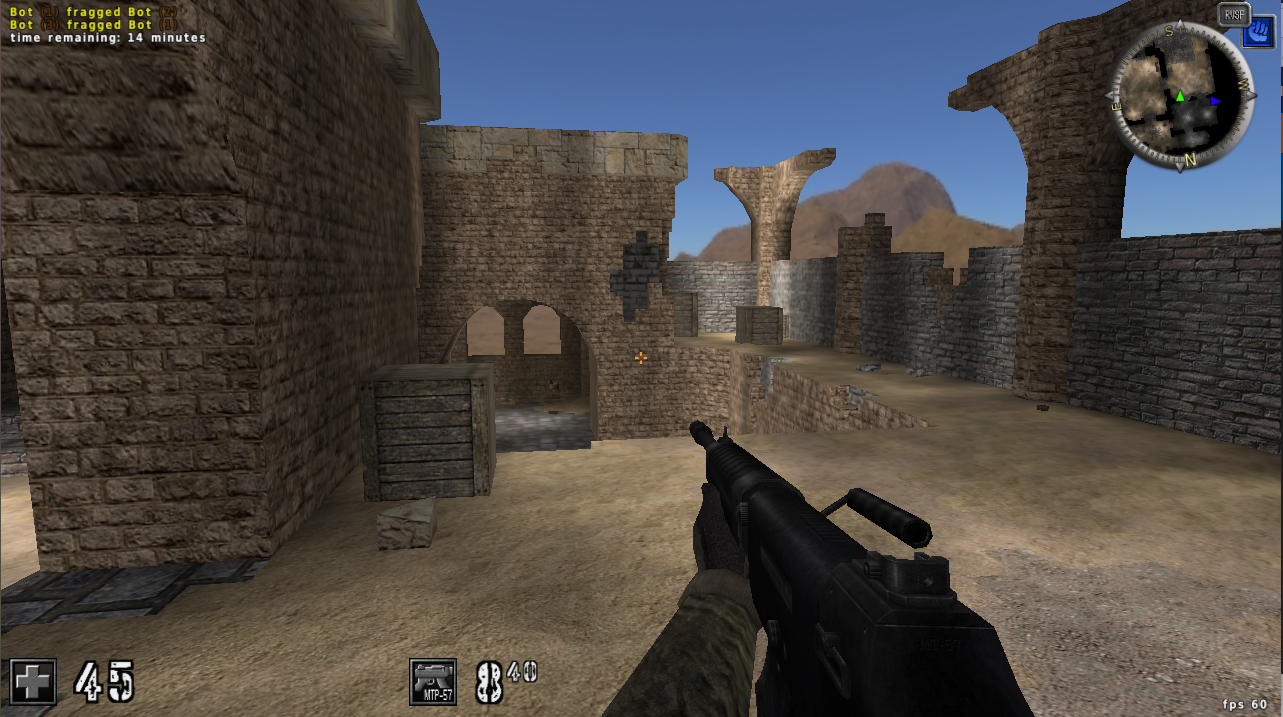}}
\subfloat[Little Racers STREET]{\includegraphics[width=0.5\linewidth]{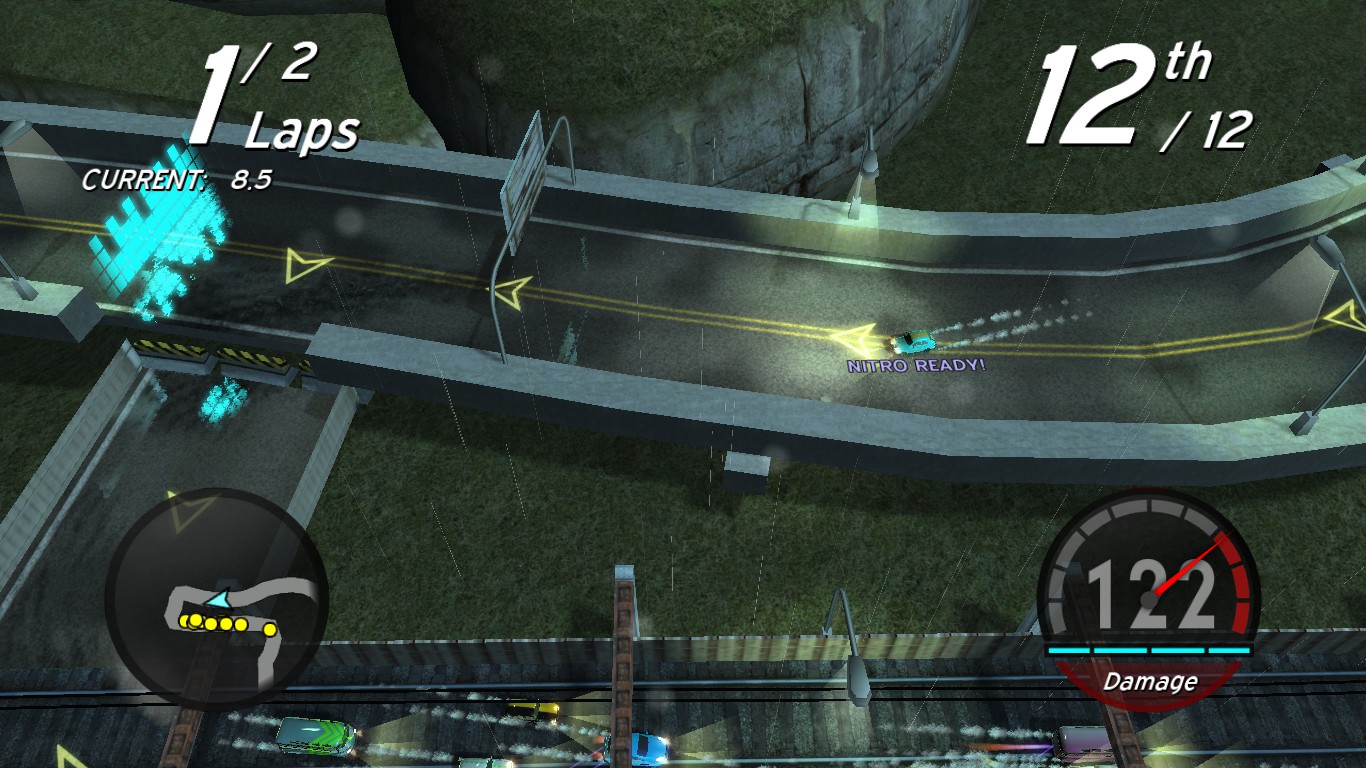}}\\
\subfloat[Formula Fusion]{\includegraphics[width=0.5\linewidth]{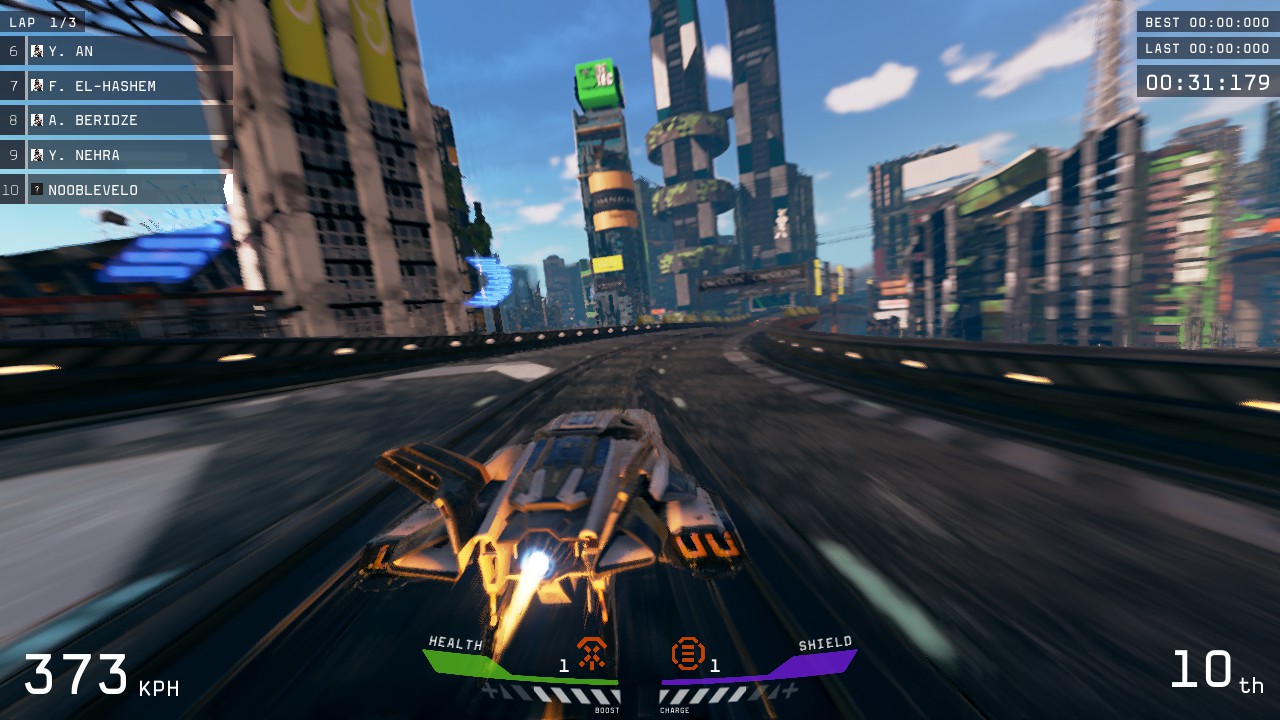}}
\subfloat[Trine 2]{\includegraphics[width=0.5\linewidth]{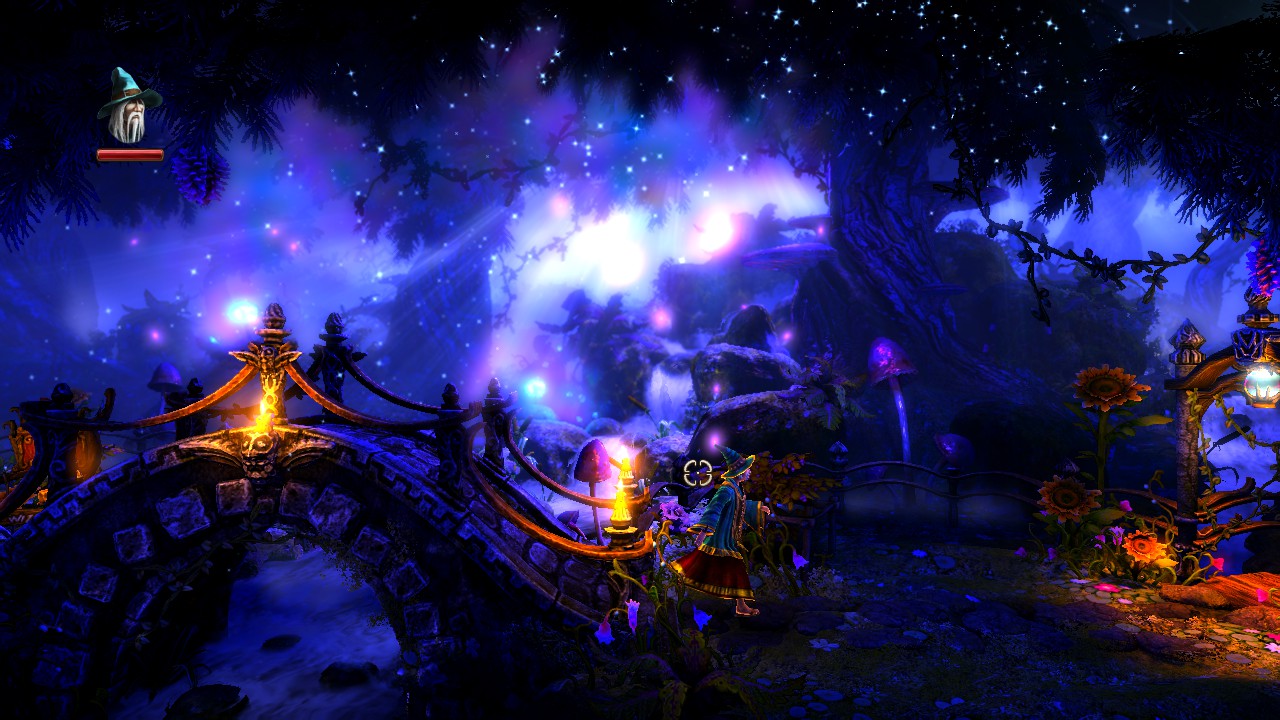}}
\caption{Sample screen captures of the games whose gaze patterns are considered}
\label{fig:screencaps}
\end{center}
\end{figure}
\begin{figure}[t]
\begin{center}
\subfloat[AssaultCube]{\includegraphics[width=0.24\linewidth]{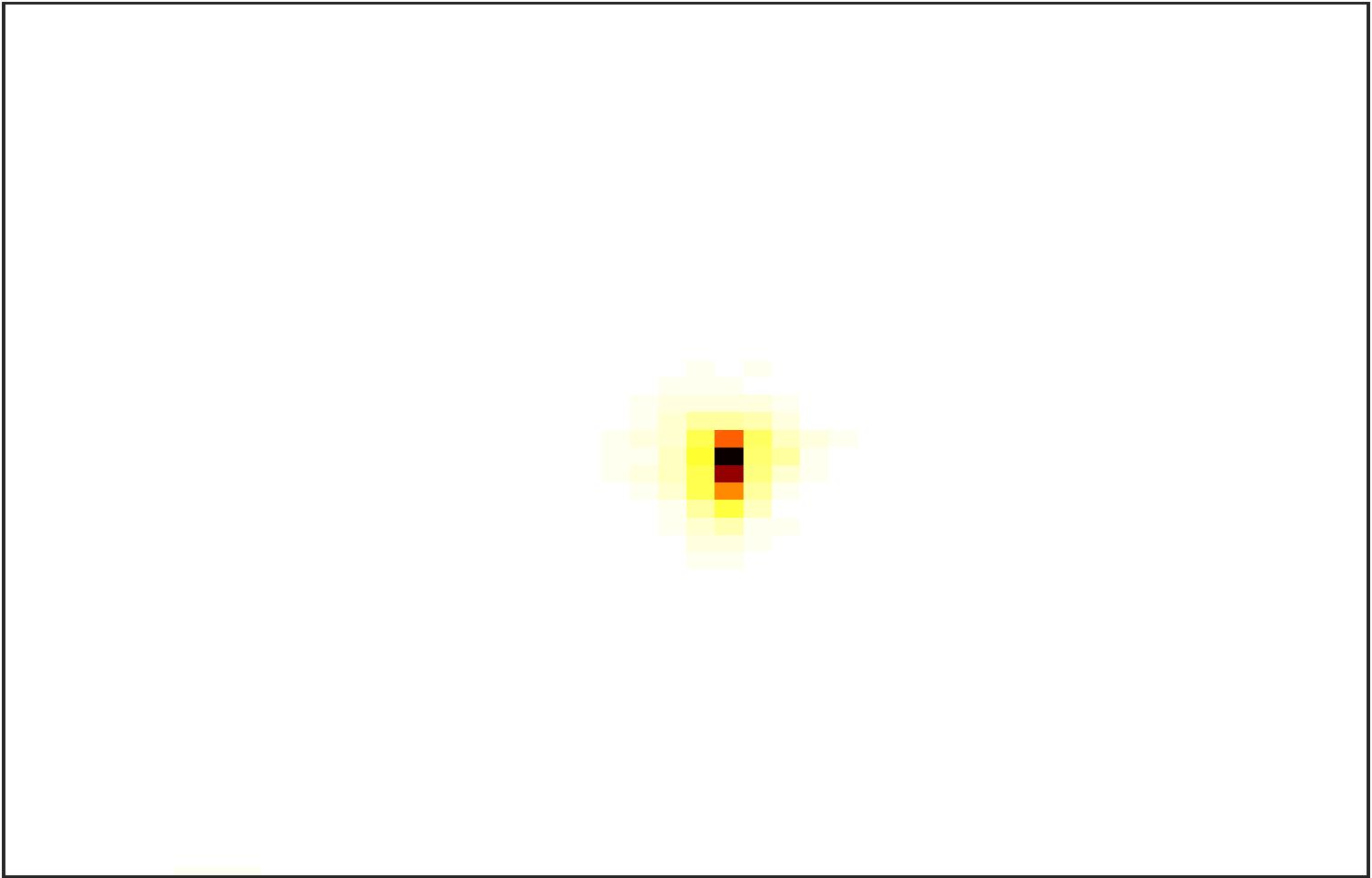}}
\subfloat[Little Racers STREET]{\includegraphics[width=0.24\linewidth]{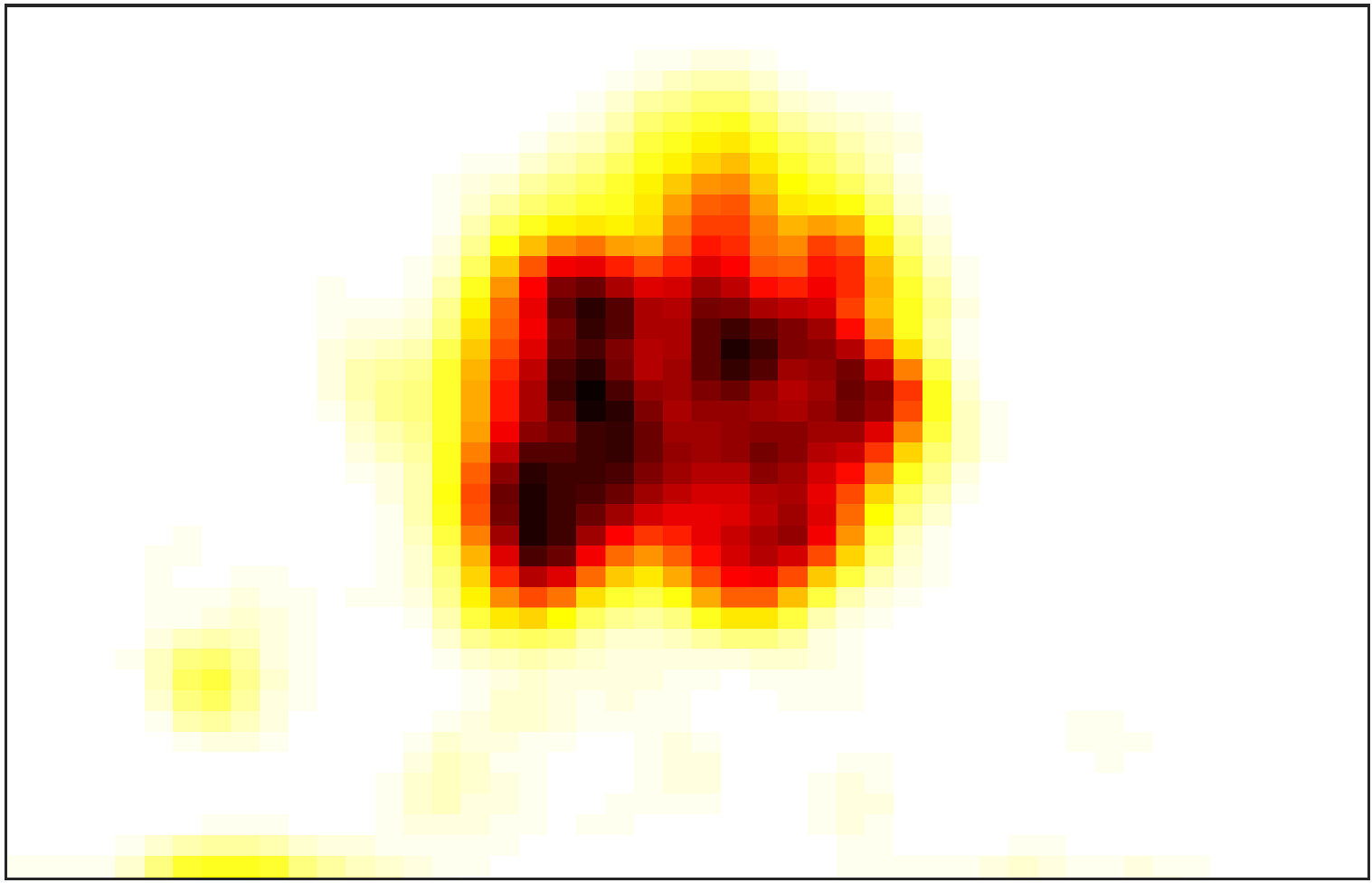}}
\subfloat[Formula Fusion]{\includegraphics[width=0.24\linewidth]{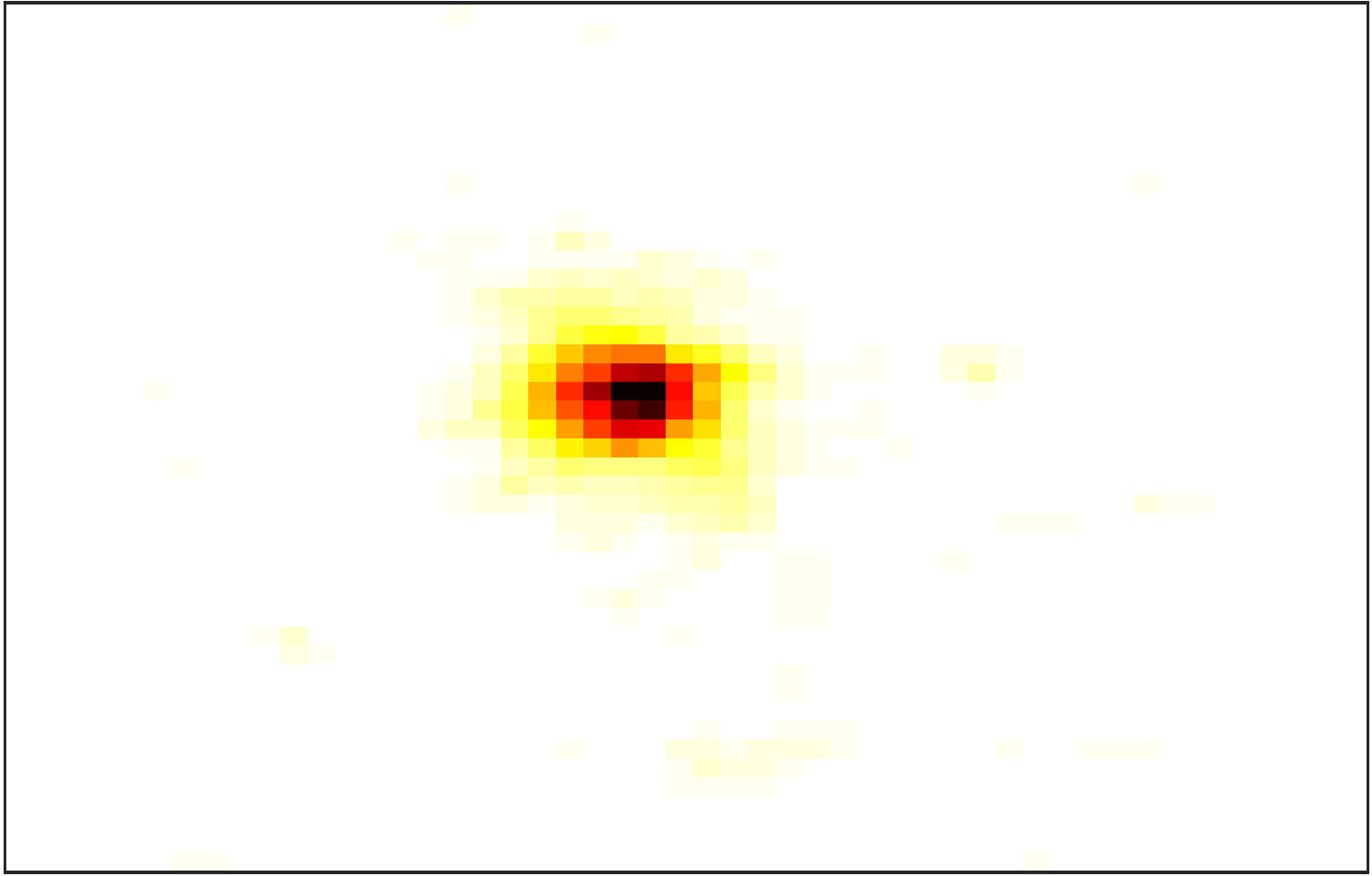}}
\subfloat[Trine 2]{\includegraphics[width=0.25\linewidth]{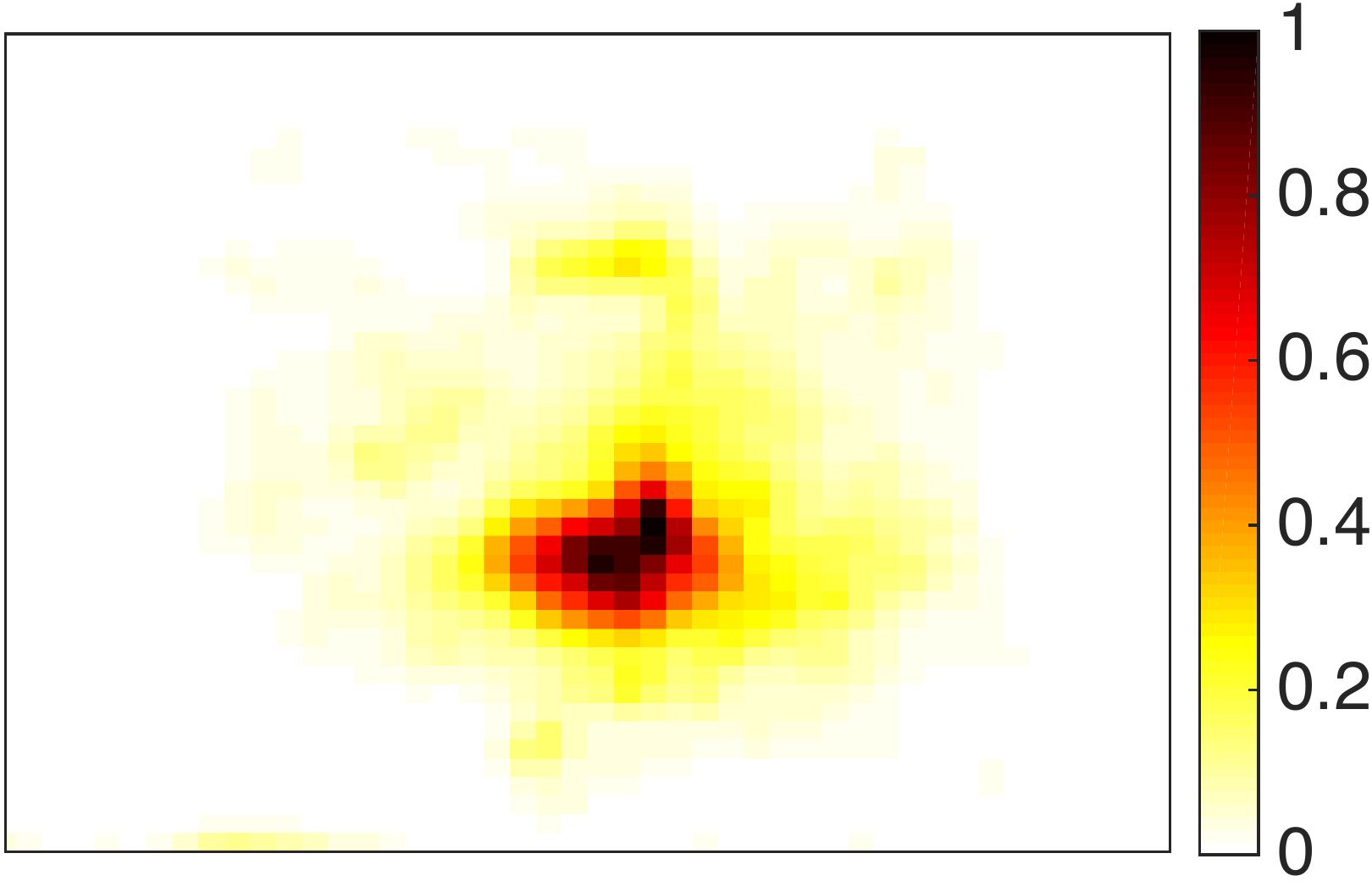}}
\caption{Gaze tracking heatmaps from 15 minute gameplay sessions. The color scale is normalized.}
\label{fig:gaze_heatmaps}
\end{center}
\end{figure}

The gaze data for each game is plotted as heatmaps in Figure \ref{fig:gaze_heatmaps}, using bivariate Gaussian kernel density estimation of gaze coordinates. Inspecting the heatmaps, the highest density of gaze coordinates, for all games, is at the center of the screen. This is expected as most gameplay graphics recenter at or around the center of the frame. In AssaultCube, which is an FPS game, the gaze coordinates are highly localized to the center of the frame. In FPS games, the player's attention is directed towards the cross-hairs of their weapon most of the time, which is usually located in the center of the frame. There might be occasional glances to various information icons, like the map and in game incident reporter (see Figure \ref{fig:screencaps}, but they are too few in number to register in the heatmap). The gaze location heatmap of Little Racers shows the widest spread of gaze location around the center. Again this is explained by the nature of the game play; the player's point of attention is the car they control around a race track. The race track although occupying the whole frame, is re-centered as the car is moved, to and around the frame center. Gaze location data for Trine2, where the player controls an avatar from a two dimensional side view, shows wider spread of gaze locations than AssaultCube or Formula Fusion. The avatar can move at least forward (right), backward (left), and up (jump), explaining the spread. Formula Fusion shows a gaze pattern similar to, but more spread out than AssaultCube, due to the fact that the car is roughly located at the center of the frame most of the time and the player may glance around to explore the upcoming track and other vehicles on the track. Glances to the periphery of gameplay frame are too low in number to register in the heat map for all games except Little Racers where some gaze locations at the bottom left of the frame register. This is due to the game play race map being located there. It is evident that the seamlessness of foveated encoding in cloud gaming depends upon the type of game play, but it may be possible to draw genre specific conclusions.

\subsubsection{Latency Considerations}
\begin{figure}[t]
	\begin{minipage}[t]{0.32\textwidth}
		\includegraphics[width=\linewidth]{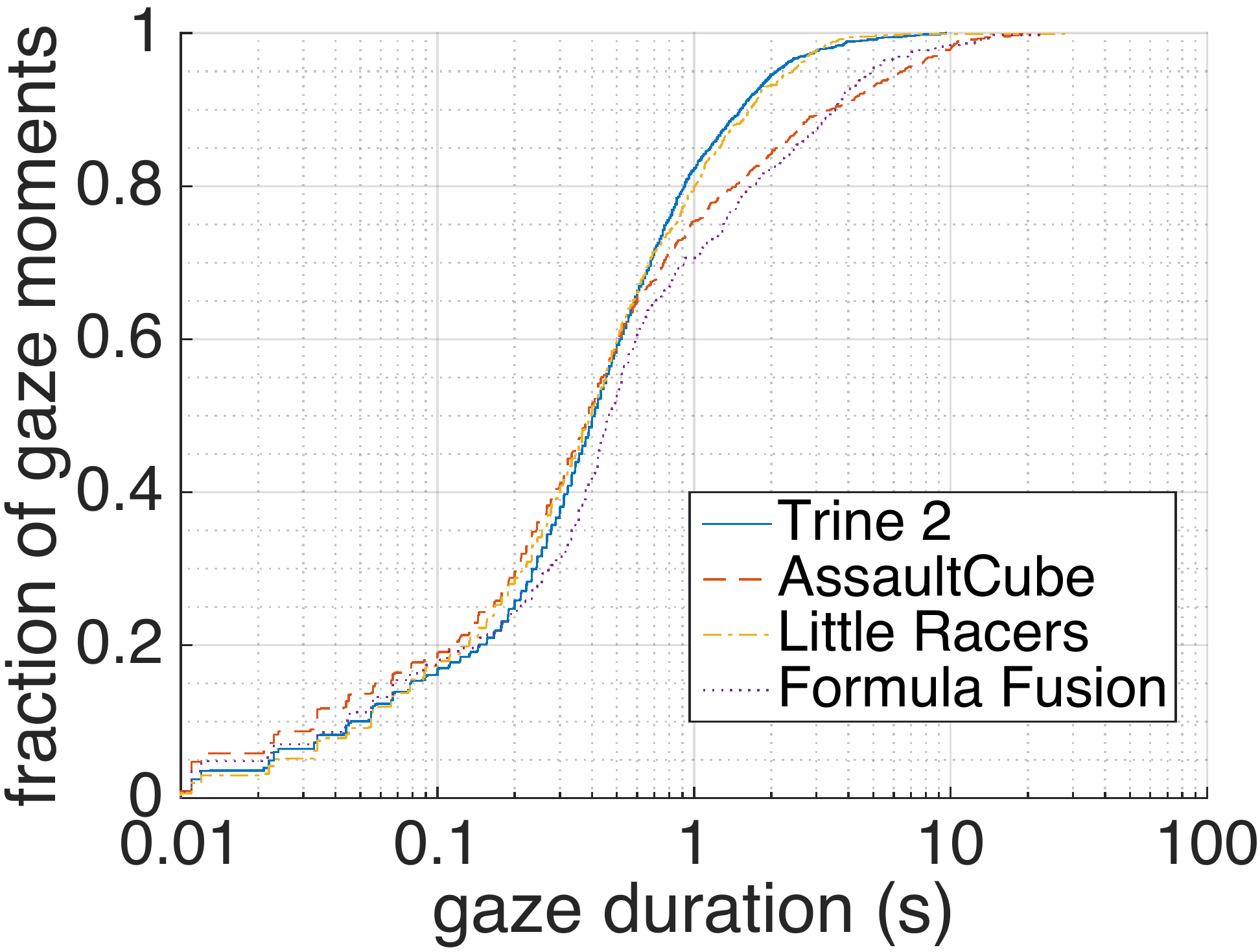}
		\caption{CDF of gaze moment duration when $W=FW/8$.}
		\label{fig:gaze_durations1}
	\end{minipage}
    \hfill
	\begin{minipage}[t]{0.32\textwidth}
		\includegraphics[width=\linewidth]{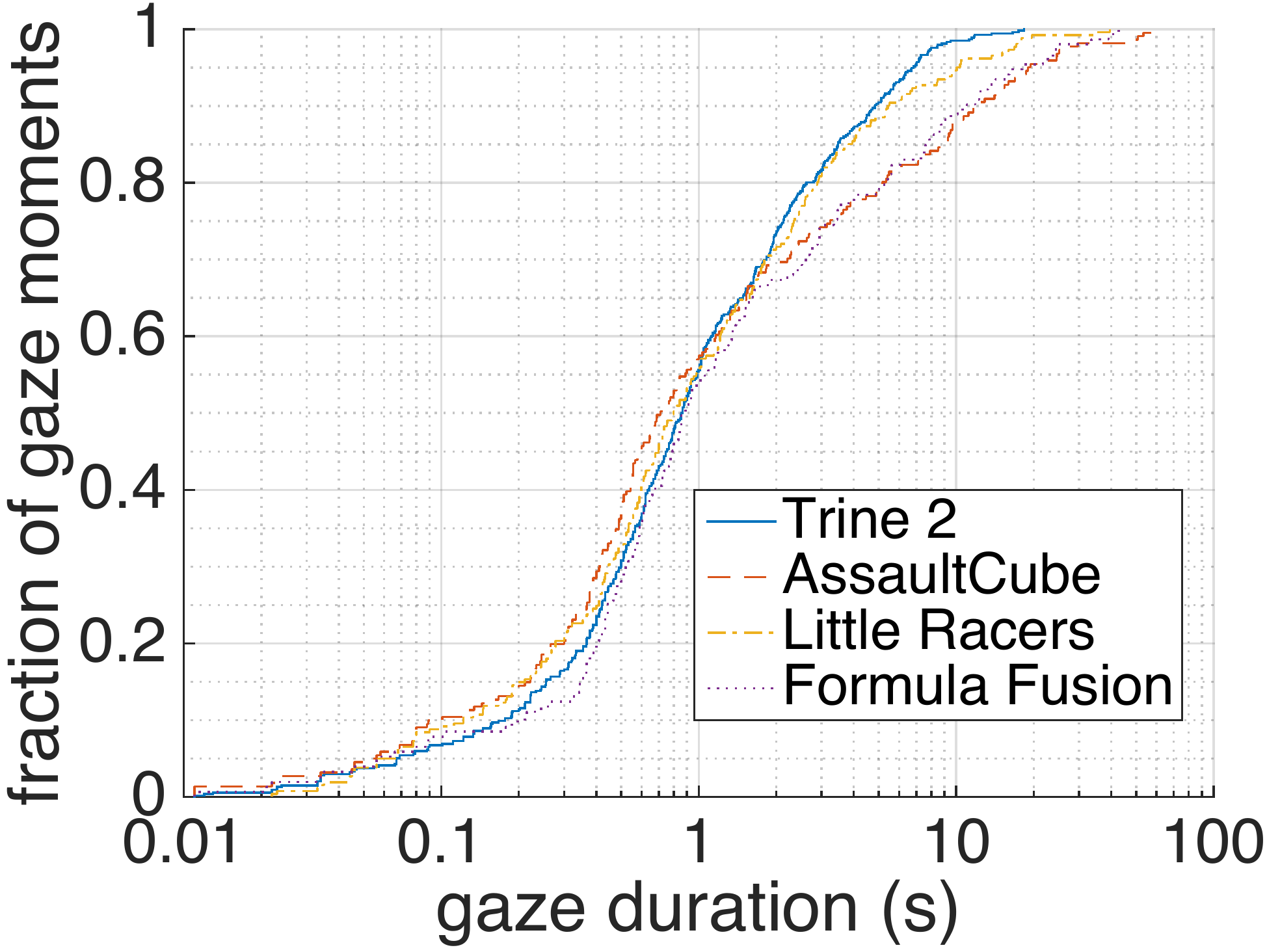}
		\caption{CDF of gaze moment duration when $W=FW/4$.}
		\label{fig:gaze_durations2}
	\end{minipage}
    \hfill
	\begin{minipage}[t]{0.32\textwidth}
		\includegraphics[width=\linewidth]{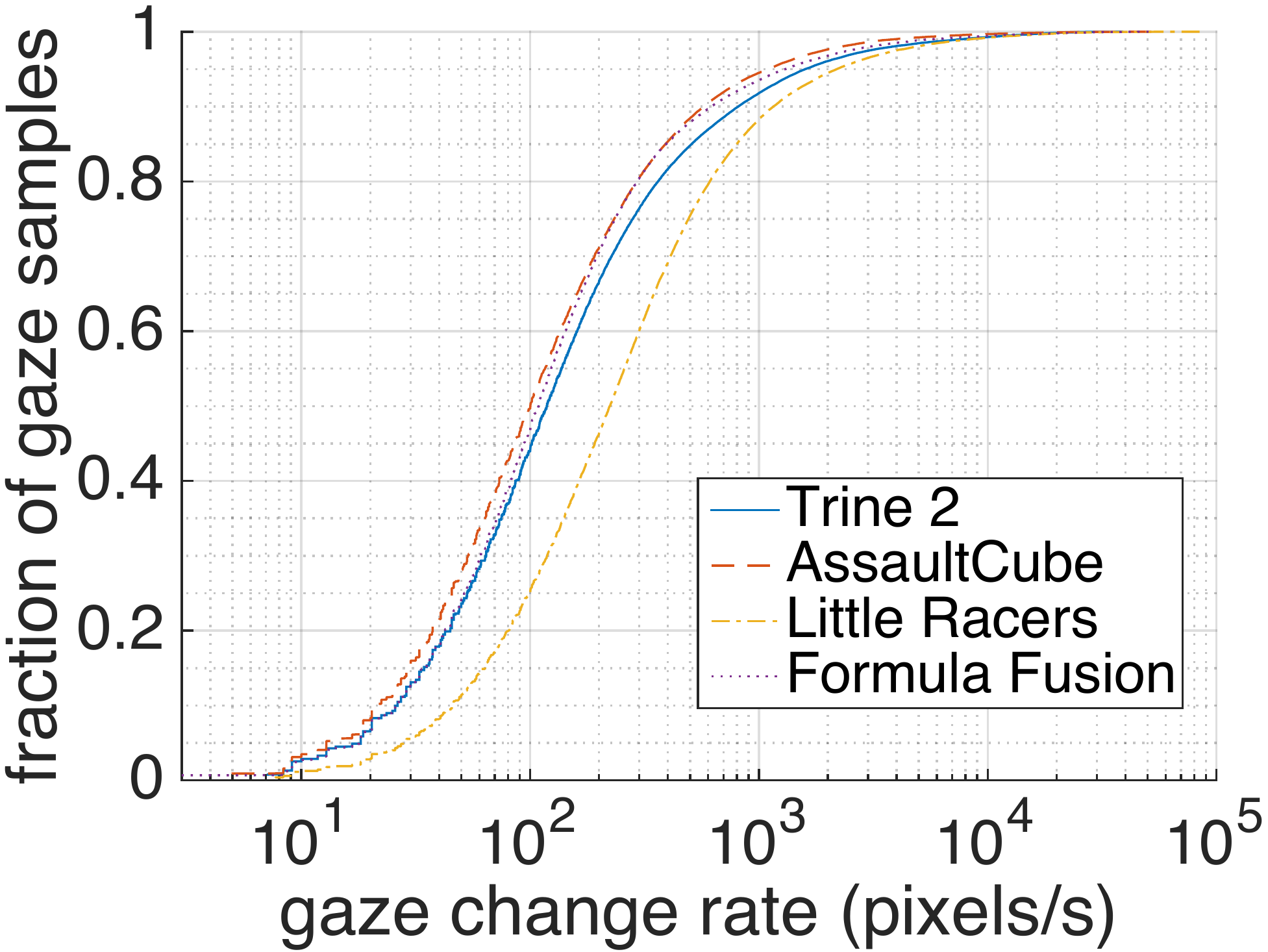}
		\caption{CDF of rate of gaze shifting.}
        \label{fig:gaze_changerate}
	\end{minipage}
\end{figure}

We next examine \textit{gaze moments} which we define as time periods within which the user's gaze lingers within a circular region of a certain radius. We define regions of two radii, $FW/8$ and $FW/4$ whose cumulative distribution functions of gaze moments are plotted in \ref{fig:gaze_durations1} and \ref{fig:gaze_durations2} respectively. From the plots it is observed that the user's gaze lingers within a region of radius $FW/4$ and even $FW/8$ almost all the time for a time slot longer than the sampling frequency of the eye tracker (which is approximately 10ms). The scenarios where gaze moments are shorter than 10ms are most challenging in terms of providing a seamless gaming experience without the user observing foveation, considering the end to end latency. Further, note from the plots in either definition of the gaze region ($FW/8$ or $FW/4$), a vast majority of the gaze moments, about 80-90\%, last longer than 100ms for all the games. Long gaze moments which last longer than 1s comprise about 20-40\% of the gaze moments, expectedly for AssaultCube and Formula Fusion these long gaze moments comprise a larger fraction. 

To investigate how fast a player's game moves while playing the considered games, we compute the rate of change of gaze during gameplay. Rate of change of gaze is calculated by dividing distance of consecutive gaze data samples, in pixels, by the time difference of samples \footnote{It should be noted that the Tobii 4C eye tracker uses some filtering based on previous gaze location and age of the sample, this might affect the results. The filtering algorithm is not in public domain and hence we do not know how much the effect is.}.  Figure \ref{fig:gaze_changerate} illustrates the CDF of the results. 
It is evident that rate of gaze change beyond 1000 pixels/s, which indicates across screen glances, are rare. Even at 1000 pixels/second, at 40-45 fps encoding used in our experiments, the per frame change in gaze location is less than 25 pixels. For the vast majority of gaze changes which exhibit slower rate of change of gaze, it follows, the per frame change in gaze is even smaller.


Latency in cloud gaming has been the focus of research by many researchers \cite{Chen2011} ,\cite{chen2014tmm}, \cite{Choy2012},including us \cite{kamarainen17mmsys}. Previous work on quality of experience of cloud gaming suggests a latency threshold of 100ms \cite{Jarschel_Qoe} beyond which the player QoE begins to degrade. In \cite{kamarainen17mmsys}, achievable end to end latency (E2E) between user control action to corresponding change in video frame at the client is investigated. It is observed that with a low enough network latency of 20-30ms, and well provisioned compute and render resources, a sub 100ms E2E is achievable. Recent work on latency requirements in VR \cite{Albert2017FoveatedVR} reports an eye to image latency figure between 50ms to 70ms for seamless foveated rendering \footnote{It should be noted in VR rendering scenarios, rendering latency plays a much more important and complicated role compared to desktop rendering. This is because of certain physiological effects of VR environments collectively called VR sickness.}. However, the authors of \cite{Albert2017FoveatedVR} note that results are conservative as the test subjects were specifically asked to  look for artifacts in the peripheral regions. In a natural gaming environment or even video viewing conditions, the latency requirements may be even less stringent. Further, with increase in the size of foveal region (which the authors parametrize as a combination of eccentricity and blur radius) higher values of latency may be bearable.
Characterizing the latency of the Tobii Eye Tracker 4C is beyond the scope of this work. However, the latency of Tobii EyeX, the predecessor to Eye Tracker 4C has been found to be about 50 ms \cite{Gibaldi2017}. Tobii Eye Tracker 4C has a higher sampling rate than Tobii EyeX and uses purportedly improved hardware and algorithms, it may be assumed that the latency of the Tobii eye tracker is less than 50ms as well. Considering this latency in the e2e latency of a modern (mobile) device in a cloud gaming setup, by replacing the device-to-kernel latency of the device with the eye tracker latency \cite{kamarainen17mmsys}, a sub 100ms  e2e latency is still possible. Now considering Figures \ref{fig:gaze_durations1} and \ref{fig:gaze_durations2}, majority of gaze moments last longer than 100ms and hence longer than the time it takes to update the foveal region. Further, from Figure \ref{fig:gaze_changerate}, for all games except Little Racers, for 50\% of the time, the gaze changes by less than 10 pixels from the time the eye gaze is located to the time the foveated region is updated on the screen. With suitably defined values for $W$ parameter, foveated encoding should be transparent to the user. However, since QoE is a highly subjective experience, we conduct a user study to validate these postulations and inferences.

%% file: user_study.tex
\section{User Study}
\label{sec:user_study}
\subsection{Method}
 \label{sec:us_method}
In order to determine what parameterizations of the foveated video encoding (FVE) are optimal with regards to QoE and bandwidth usage, we set up a controlled laboratory experiment to gather data on the perceived video quality and player's experience in using our foveated cloud gaming system. In the experiment, participants played a game on the prototype with different FVE parameters. The goal of the experiment was twofold: Firstly, to determine whether users notice if foveated video encoding is being used, and, if so, how strongly FVE influences their experience in playing a game. Secondly: to compare the subjective assessments of the video quality to the bandwidth usage for several different parameterizations to try and determine a relationship between them. Using the latter comparison we show support for the hypothesis that there is an optimal parameterization with regards to bandwidth usage and video quality.
\begin{table}[h]
\caption{Dependent variables recorded in the experiment per session, using 100-point Likert scales. Before the experiment started the participant was explained what was meant by each question.}
\label{tab:depvars}
\begin{minipage}{\columnwidth}
\begin{center}
\begin{tabular}{ll}
\textbf{Variable}	& \textbf{Explanation}\\
\toprule
Video quality	& Quality of the video image (pixelation, artefacts, etc. as opposed to detail of in-game models)\\
Video adequacy	& How adequate the video was for their task, i.e. did the video hinder their performance or not?\\
Enjoyment	&	How enjoyable was the experience?\\
Satisfaction	& How satisfied are you with your performance (score)?\\
Effort	& How much effort did you put into completing the task?\\
Concentration	& How well were you able to keep your concentration?\\
\bottomrule
\end{tabular}
\end{center}
\end{minipage}
\end{table}
\subsection{Participants}
We recruited 12 participants from the Computer Science building at Aalto University. In order to make sure that their ability to play the game itself was not a confounding factor, we invited only participants that had some experience in playing First Person Shooter (FPS) games on PC (i.e. using a keyboard and mouse as the controller).
The participants were aware only of their task, the data we collected and that the system was based on cloud gaming. Only after the experiment was finished were the participants told of the purpose of the experiment and the particular kind of foveated encoding technology that was used.
\subsection{Experiment Design}
The experiment was a within-subjects design with the maximum quantization parameter ($QO_{max}$) as the independent variable. The independent variable was divided into 5 levels with equal intervals from highest to lowest maximum quantization offset, where a $QO_{max}$ of 16 to the maximum level of quantization and $QO_{max}$ of 0 corresponds to no foveated encoding.
The dependent variables were video quality, video adequacy, enjoyment, performance satisfaction, effort and concentration which were measured through respective 100-point Likert scales (see Table \ref{tab:depvars}. The remaining dependent variables were the gaze data (output from the client software), the bandwidth and (measured through Wireshark on the client), participant comments (taken after each session), and the participant's score.

The rating scale we used is adapted from Mullin et al.\cite{Mullin2001} and is further inspired by Pauliks et al.'s work\cite{Pauliks2013}. Pauliks et al. argue that for short video presentations any method of assessing video quality is equally well-suited. Mullin et al. discuss the problems with ITU recommended video quality assessment scales, and propose the use of a 0 - 100 rating scale without labels. In this study, we asked the participants to rate the video quality and how adequate the video was for the task on a 0-100 rating scale. The other questions related to the participant's enjoyment, performance, concentration and effort, respectively. For an overview of the rating scale questions and their explanation as given to the participant, see Table \ref{tab:depvars}.

\begin{figure}[ht] 
 \begin{center} 
\includegraphics[width=.6\columnwidth]{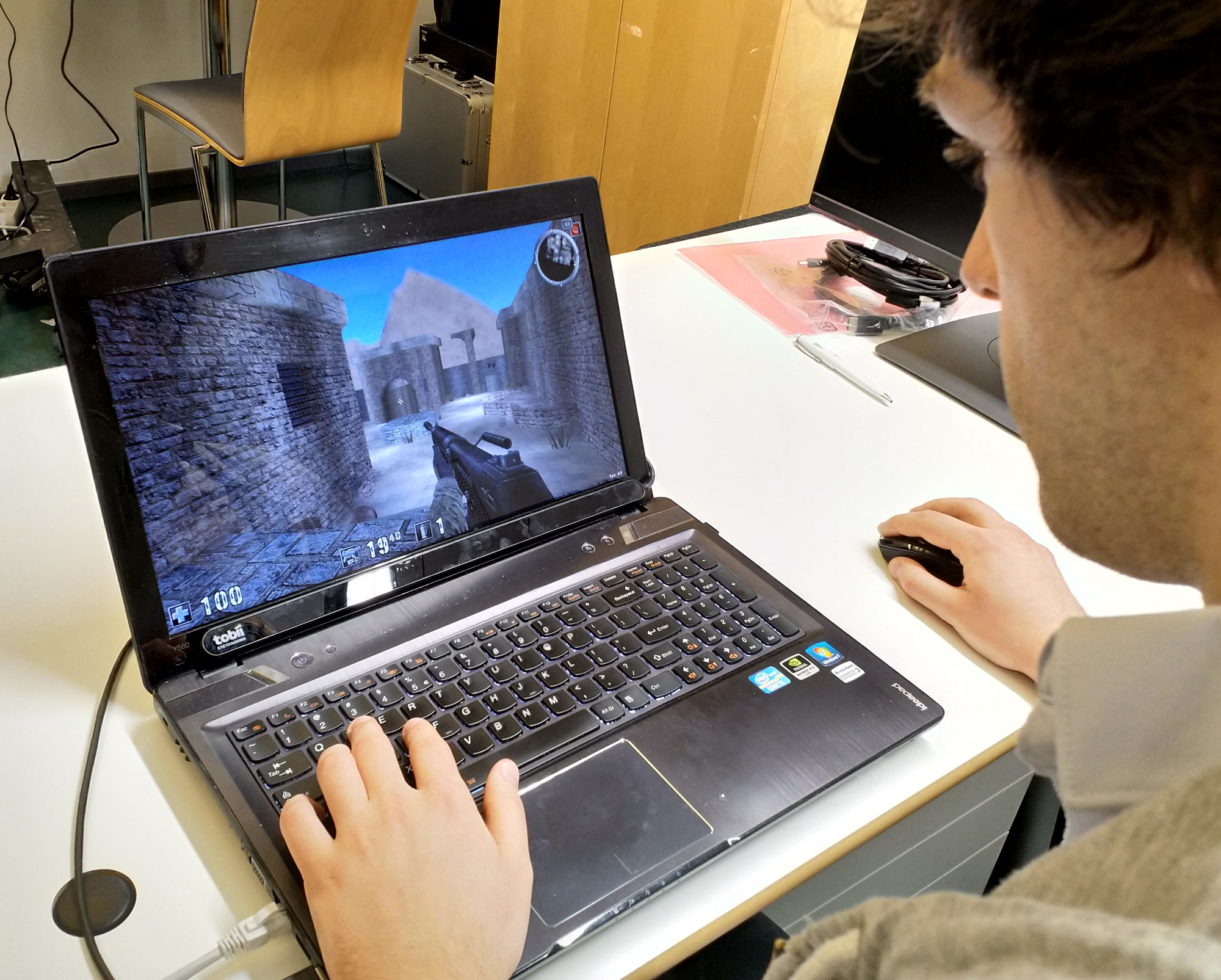} 
\caption{\small \sl Experimental setup showing the Tobii 4C eye-tracker (bottom of the screen), peripherals and the Lenovo Y580 client device rendering AssaultCube on the Desert map. \label{fig:setup}} 
 \end{center}
 \vspace{-5mm}
 \end{figure}

The participant was seated about 50 cm from the screen of the client device, with the eye-tracker attached to the lower part of the screen and aimed towards the user's face. The set-up can be seen in Figure \ref{fig:setup}: the client device is connected using a wired ethernet connection, and the participant plays the game using a wireless mouse and the laptop's keyboard. For further details, see the following subsection on Apparatus (\ref{sec:apparatus}).

The foveated area was set as a circle with a radius of 1/8th of the screen size: recall from previous sections that the foveated area of the human visual system is about 2 degrees, and that the effective width depends on the distance between the participant's eye and the screen. Given the distance to the screen of about 50 cm and a screen width of 40 cm, we calculate the foveated area to be about 2 cm, which is about 1/20th of the screen width. The W parameter was set conservatively to $W = FW/8$ to ameliorate sudden gaze movements and any inaccuracies in eye tracker data.

\subsection{Materials \& Apparatus}
\label{sec:apparatus}
The client and server of the prototype system as described in \ref{sec:system} are deployed respectively on a Lenovo IdeaPad Y580 laptop with a 15.6" full-HD screen running Windows 10 64-bit and an Ubuntu Linux 16.04 LTS 64-bit host with a Nvidia GeForce GTX 1050Ti graphics card. The game used is Assault Cube v1.1.0.4 and the map used is 'Desert' with 7 bot players in a Free-for-all game mode ('Bot Deathmatch'). A Tobii 4C eye-tracker was attached to the lower part of the laptop screen, aimed towards the user's face and calibrated for each participant. The participant was seated on a general-issue chair at a fixed distance from the laptop and the laptop and participant are aligned on the same axis (Figure \ref{fig:setup}). The participant used a standard issue wireless optical mouse to control the game, together with the laptop's keyboard. The bandwidth data was recorded using Wireshark 2.6.0, while the client device and server were connected on different subnets of the campus intranet on a 1 Gbps wired connection.

\subsection{Procedure}
For each participant the experiment consisted of five sessions of 5 minutes each, where each session used a different $QO_{max}$ setting. The maximum quantization offsets we used were $QO_{max} = 0, 4, 8, 12, 16$. The starting condition was rotated, while the order of the different conditions was fixed in the following sequence: $QO_{max}$ =  8, 4, 12, 0, 16 for the respective sessions. The task, for each session, was to get at least 40 kills (amount of deaths was mentioned to be irrelevant). Before the experiment the participant played a warm-up game (at $QO_{max} = 0$) for 2 minutes, in order to get comfortable with the game controls and to ensure the functioning of the software.

\textbf{(1)} The participant was invited in, welcomed and explained the general idea of the experiment and the procedure. Before starting the experiment the participant was also explained in basic terms what constitutes video quality (as opposed to in-game graphics quality). The participant was not told about the foveation or encoding technology used in the system, to avoid hypothesis guessing. We explained the questions on the rating scales and asked the participant to read and sign the consent form.

\textbf{(2)} The participant was then placed on the chair in front of the laptop, and asked to sit comfortably in front of the computer considering the distance and alignment. The eye-tracker was calibrated and the warm-up game started, during which the controls of the game and the task were explained. Using the self-reported skill level of the participant and their performance in the warm-up game, the difficulty level to use was determined to be 'Worse', 'Medium' or 'Good'. 

\textbf{(3)} After the warm-up game, and after answering any questions from the participant and confirming that the participant was ready, the first condition of $QO_{max}$ was started. When the participant commenced the task, we started a timer for 5 minutes, after which we determined and recorded the participants score in the game and asked the participant to rate enjoyment, performance satisfaction, video quality, video adequacy, effort and concentration on respective 100-point Likert scales. We also asked the participant to comment in general on their thoughts about playing that session. This step was repeated for subsequent conditions with the different respective settings of $QO_{max}$.

\textbf{(4)} After the last session was finished as per step 3, we explained to the participant what technology we were using (particularly the foveated video encoding) en the purpose of the experiment. We then asked them to, with this new knowledge, comment on their experience.

\begin{table}
\caption{(Parameter) settings used in the experimental setup.}
\label{tab:vars}
\begin{minipage}{\columnwidth}
\begin{center}
\begin{tabular}{lll}
\textbf{Parameter}	& \textbf{Setting}	& \textbf{Note}\\
\toprule
$QO_{max}$	& 8, 4, 12, 0, 16	& Varied per subsequent session, in this order. Starting condition rotates.\\
$W$	&	8	& Fixed, used to control the size of the foveated area.\\
Render resolution	& 1920 x 1080	& Resolution set in the game and fed to the encoder in pixels, fixed.\\
Video resolution	& 1920 x 1080	& Resolution of the video stream the client receives in pixels, fixed.\\
Video FPS	& 50	& Frames-per-second set in the encoder.\\
Graphics quality	& Highest	& Relates to the video quality settings built into AssaultCube.\\
Goal score	& 40 kills	& Target score in kills the participant needed to reach, deaths was irrelevant.\\
Difficulty setting	& W(1), M(6), G(5)	& Bot difficulty: Worse, Medium, Good with number of assignments.\\
Wireshark filter	&	& \textit{host [Server IP] and not port 22 and not port 5900}\\
Eye-tracker	&	& Calibrated at the start of the experiment for each participant.\\
\bottomrule
\end{tabular}
\end{center}
\end{minipage}
\end{table}
\begin{figure}
\begin{minipage}[t]{0.49\textwidth}
\begin{center} 
\includegraphics[width=\textwidth]{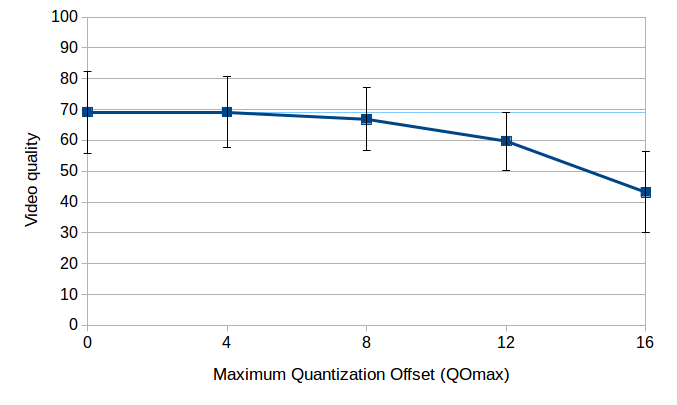} 
\caption{\small \sl MOS for the video quality (scale 0 - 100), by maximum quantization offset with 95\% Confidence Interval. The light blue line at video quality = 69 represents the MOS for $QO_{max} = 0$. \label{fig:video1}} 
\end{center} 
\end{minipage}
\hfill
\begin{minipage}[t]{0.49\textwidth}
\begin{center} 
\includegraphics[width=\textwidth]{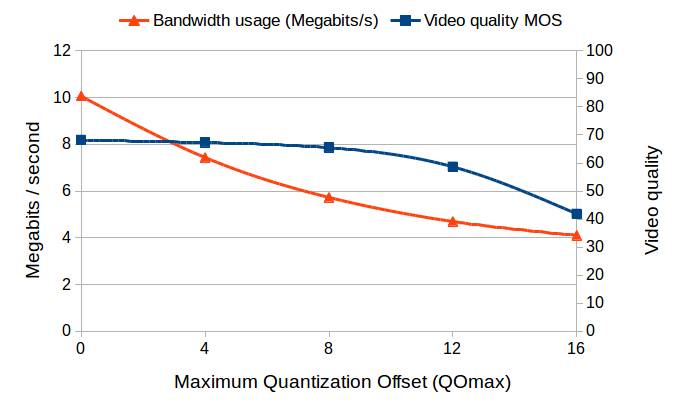} 
\caption{\small \sl MOS for the video quality (scale 0 - 100) plotted against the bandwidth usage in average megabits per second, by maximum quantization offset.
\label{fig:video_bandwidth}} 
\end{center} 
\end{minipage}
\end{figure}
\begin{figure}
\begin{center}
\includegraphics[width=0.5\textwidth]{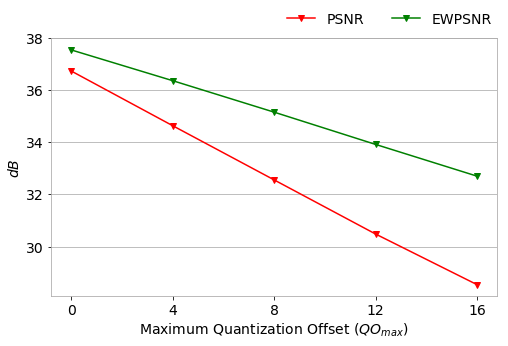}
\caption{PSNR and EWPSNR values of foveated gameplay video sequence at different $QO_{max}$ values. The gaze was fixated at the center of the frame during foveated encoding}
\label{fig:ewpsnr}
\end{center}
\end{figure}

\subsection{Results: Foveation and QoE}
\label{sec:us_discuss}
In this section we present and discuss the results of the user study. We first study the Mean Opinion Scores (MOS) for video quality and objective evaluation of video quality, followed by the results of the bandwidth usage logging and the reported scores for the other questions on the rating scale that served as a control and relate more to the task overall. We aggregated the scores per participant and grouped them by maximum quantization offset ($QO_{max}$ parameter). Recall that the $QO_{max}$ setting was varied per condition, so the 5 different levels correspond directly to the five different sessions and their respective conditions. The score range on all rating scales was 0 - 100, where 0 is the lowest score, and 100 the highest. Participants were free to give any rating on this scale. We calculated the Mean Opinion Score as the mean of the responses per maximum quantization offset.\footnote{Since our main interest is in the difference in scores, we wanted to give participants as much freedom as possible to rate the respective variables. This method is adapted from \cite{Mullin2001} and is further inspired by \cite{Pauliks2013}.} We used a Student's t-distribution to estimate the mean and calculate the 95\% confidence intervals, due to the relatively low number of samples (12). After a participant had finished their sessions, we asked for free form comments and also informed them about the purpose of the experiment asking them whether they had noticed FVE.

In order to understand how the results compare to those obtainable with objective video quality metrics, we also calculate PSNR and Eye Weighted PSNR (EWPSNR)\cite{Li_2011} of a foveated encoded gameplay video sequence at various $QO_{max}$ values. The foveated encoding considered the gaze to be fixated at the center of the frame.

\subsubsection{Video Quality}
In Figure \ref{fig:video1} maximum quantization offset ($QO_{max}$) is plotted together with the Mean Opinion Score of the video quality.  In Figure \ref{fig:video_bandwidth}, the video quality MOS and bandwidth usage are plotted against the maximum quantization offset. Note that the left y-axis (average megabits per second) relates to the bandwidth usage and the right y-axis (score 0 - 100) to the video quality MOS. In Figure \ref{fig:video1} we can see a drop-off in perceived quality when $QO_{max} > 8$. We can see that $QO_{max} = 4$ is rated equal to using no foveated encoding ($QO_{max} = 0$), and that at $QO_{max} >= 12$ users consistently rate the quality to be low. This corresponds to the free-form comments we received during the experiment, where 7 out of 12 participants commented on the video quality being much worse suddenly for $QO_{max} = 16$, and 3 out of 12 for $QO_{max} = 12$. The bandwidth usage as depicted in Figure \ref{fig:video_bandwidth} follows that of results in Section \ref{sec:Fov_throughput}, showing a logarithmic decrease in bandwidth usage with maximum quantization offset. The video quality is best at $QO_{max} = 0$, and consequently requires the most bandwidth. At about $QO_{max} = 8$ the bandwidth usage starts to plateau. Considering both video quality and bandwidth usage, it is clear that between $QO_{max}=8$ and $QO_{max}=12$, there is scope for finding a sweet spot where the bandwidth savings are significant while QoE is minimally affected.

The average PSNR and EWPSNR for different values of $QO_{max}$ are shown in Figure \ref{fig:ewpsnr}. It is clear that with increase in $QO_{max}$ both PSNR and EWPSNR become smaller, but the drop in PSNR values is steeper, which suggests that EWPSNR indeed accounts for the foveated encoding to some extent. At $QO_{max}$ = 8, the drop in EWPSNR is just 2dB and at $QO_{max}$ = 12 the drop is less than 4 dB indicating a presumably lower loss of quality, which is corroborated by the MOS scores in Figure \ref{fig:video1}. Interestingly, MOS scores decrease nonlinearly as a function of $QO_{max}$, while PSNR and EWPSNR values decrease linearly. It is illustrative of the fact that objective metrics, even the ones that adapt to gaze location like EWPSNR, may not precisely reflect the viewer perceived video quality when foveated video encoding is applied. Further work is needed to understand the root cause of this discrepancy and hopefully develop more appropriate metrics.

\begin{figure}
\begin{minipage}[t]{0.48\textwidth}
\begin{center}
\includegraphics[width=\columnwidth]{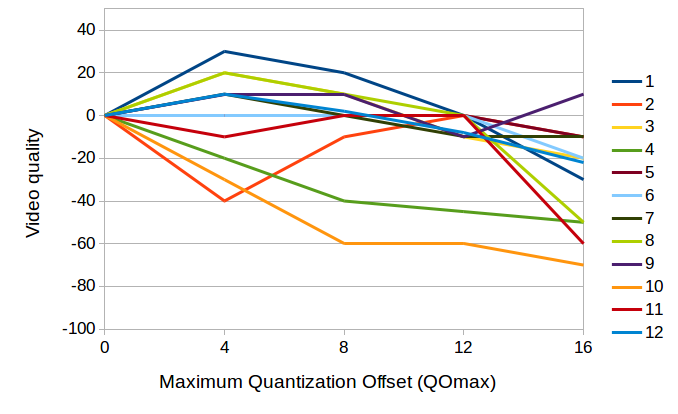} 
\caption{\small \sl Individual difference scores for the video quality per maximum quantization offset, with reference $QO_{max} = 0$.
\label{fig:indiv_mos}} 
\end{center}
\end{minipage}
\hfill
\begin{minipage}[t]{0.48\textwidth}
\begin{center}
\includegraphics[width=\textwidth]{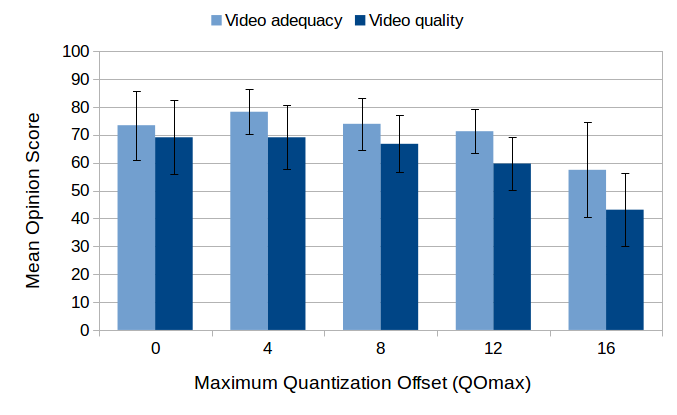} 
\caption{\small \sl Video adequacy MOS (scale 0 - 100= and Video Quality MOS (scale 0 - 100) per maximum quantization offset with 95\% confidence intervals.\label{fig:adequacy}} 
\end{center}
\end{minipage}
\end{figure}
We also plot individual MOS difference scores with respect to maximum quantization offset in Figure \ref{fig:indiv_mos}. The difference is calculated with respect to the condition $QO_{max}=0$, at which the MOS difference for a user is zero.
If we take a look at the individual difference scores in Figure \ref{fig:indiv_mos}, we see an interesting picture: two participants rate the video quality to be almost linearly worse for higher  values of $QO_{max}$ (4 and 10). Upon further inspection of the comments and their data we see that one of these participants had a high level of experience, and experienced a lot of lag during the sessions. The participant reported that their experience overall was not up to their expectations, and as a result his ratings clearly diverge from the average. The second participant may have suspected that the experiment was about the video quality in an FVE context, and thus payed more attention to it. Looking at the other responses we see that there are rather large differences in how people rate video quality: At $QO_{max} = 4$ in Figure \ref{fig:indiv_mos} we can see that no participant gave the same rating in this condition. On the other hand, at $QO_{max} = 12$ there seems to be a point of convergence, where participants rate the quality as being nearly equal to the condition without foveation ($QO_{max} = 0$).
Combined with the MOS scores from Figure \ref{fig:video1}, the data seems to suggest that users do not notice significant differences in video quality until $QO_{max} > 12$. Recall however, that the starting condition was rotated and that the order of the maximum quantization offsets is different from the order on the X-axis in Figure \ref{fig:indiv_mos}.
\subsubsection{Video Adequacy and Game Enjoyment}
We also asked participants to rate the adequacy of the video after each session and plotted their responses against their video quality ratings in Figure \ref{fig:adequacy}. Here we explained to participants that by video adequacy we mean  how much they felt that the video quality allowed them do to their task (getting 40 kills) well. Hence, if the amount of artefact's is distorting a user's vision, we expect the adequacy to be rated low, while if there are no obvious hindrances, we expect the adequacy to be rated high, regardless of other aspects related to performance and quality. Comparing the MOS for video quality and adequacy of the video in Figure \ref{fig:adequacy} shows no big surprises, with most people finding the quality overall to be quite adequate for the task. Only at $QO_{max} = 16$ people reported that the lower quality was disturbing and less adequate (e.g. they failed to see items on the ground, such as ammunition boxes or players in a dark corner).
\begin{figure}
\begin{minipage}[t]{0.48\textwidth}
\begin{center}
\includegraphics[width=\columnwidth]{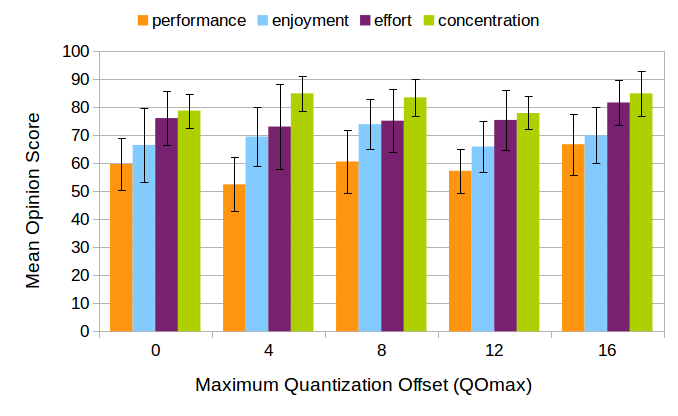} 
\caption{\small \sl Mean Opinion Scores for the task-related rating scales: Enjoyment, Performance satisfaction, Effort and Concentration. \label{fig:game_qoe}} 
\end{center}
\end{minipage}
\hfill
\begin{minipage}[t]{0.48\textwidth}
\begin{center} 
\includegraphics[width=\columnwidth]{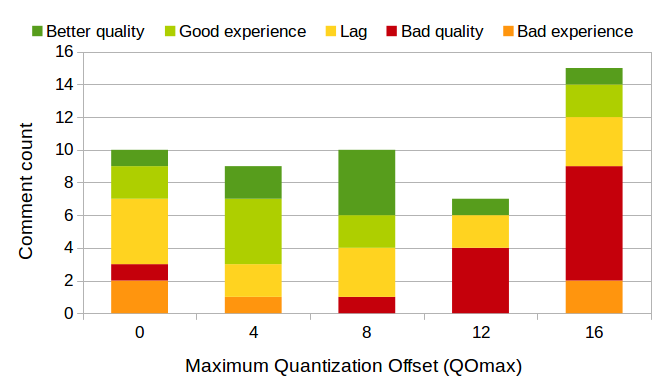} 
\caption{\small \sl Generalized comments from the participants by maximum quantization offset. Y-axis represents how many times a particular comment was given.\label{fig:comments}} 
\end{center}
\end{minipage}
\end{figure}
Finally, as a control, we asked the participants to rate how satisfied they were with their performance, how much they enjoyed the task, how much effort they put in and how well they were able to hold their concentration. Their scores are plotted in Figure \ref{fig:game_qoe}. The results are consistent over the different maximum quantization offsets, but performance satisfaction and enjoyment vary more: most likely due to different expectations and latency issues (participants with less experience reported an overall higher level of enjoyment and satisfaction regardless of score and latency issues, while more experienced players reported that their performance and possible latency issues were not up to their standards, thus lowering their respective scores on the rating scale). The consistent results may be interpreted as the players being engaged with the game not withstanding the video quality, which augments our results for video quality and adequacy.

Free form comments are roughly grouped into comment classes. In reply to whether FVE was noticeable, one participant reported that the degradation in quality (at $QO_{max} = 16$) was especially visible when he made fast gaze relocations, but otherwise did not notice this having anything to do with the use of the eye-tracker or the foveated encoding. All other participants reported to have been fully unaware of the purpose of the eye-tracker and the use of foveated video encoding or similar principles. All the participant's comments have been generalized and depicted in Figure \ref{fig:comments}. Note that the comment 'Good quality' was mainly uttered with regard to a previous session, i.e. meaning 'better quality than before'. If we look at $QO_{max} = 8$ in Figure \ref{fig:comments}, for example, we see that it has the largest count of "good/improved quality" comments. Recall from Section \ref{sec:us_method} that this condition was either played first or right after $QO_{max} = 16$, the worst quality condition. This effect is present in a similar manner in the other conditions, so we advise caution when interpreting these results.
\subsection{Limitations}
The main issue we encountered during the experiments was latency or network connection issues: during several sessions, the latency was so high it caused more or less noticeable amounts of lag in the game, as reported by the participant. In some sessions the connection dropped completely, and the session had to be restarted. This could potentially influence the participants concentration or learning effects, but we assume the effect was marginal at most, given the data and comments of the participant. 
We further suspect that hypothesis guessing may be an issue: a participant who knows about the relation between the eye-tracker and the video may (unconsciously) try to focus on their peripheral view or explicitly try to detect abnormalities. Since this behavior does not correspond to a natural way of playing a game, we tried to avoid this by giving the participant a challenging, demanding task (get at least 40 kills or more) and by not informing them of the technology used. After the experiment all of our participants reported that they were unaware of the hypothesis or the technology used during the experiment, although one participant may probably have been more informed than the others as we discussed in Section \ref{sec:us_discuss}.

Considering the large confidence intervals in Figure \ref{fig:video1} and the high variability we observed in the data, it appears that participants grade video quality very differently. This could mean that they have trouble understanding what exactly constitutes video quality, especially as opposed to graphics quality. We aimed to resolve this by recruiting participants that have some experience with video games, and explain how video quality differs from in-game graphics quality. However, given our results a more robust approach may be needed to account for individual anchoring strategies as well.

%% file: conclusion.tex
\section{Conclusions and Future Work}
\label{sec:conclusions}

In this work, we proposed to combine cloud gaming with foveated graphics. We developed a prototype system that integrates foveated streaming with off-the-shelf gaze tracker device into state-of-the-art cloud gaming software. Our evaluation results suggest that its potential to reduce bandwidth consumption is significant, as expected. We also demonstrate the impact of different parameter values on the bandwidth consumption with different games and provide some pointers on how to select parameter values. Back of the envelope latency estimations based on related work and gaze tracker specifications combined with gaze data analysis give us reason to be relatively optimistic about the impact on user experience. A  user study establishes the feasibility of FVE for FPS games, which are the most demanding latency wise. The user study underlines the significant bandwidth savings that can be accrued with suitable parameterization of FVE without sacrificing QoE. Since FPS games have significantly tighter latency constraints, we are optimistic that foveated graphics for games of other genres will show even better results.
As future work, we are planning to examine the QoE dimension in more depth through more subjective studies considering different genres of games as well as the impact of network latency. Further, we intend to attempt eliminating specialized hardware for eye tracking by employing web-cameras for the purpose. Using web cameras, which are ubiquitous in modern consumer computing devices like netbooks and mobile devices, would enable widespread adoption of foveated streaming for cloud gaming. We also intend to investigate the feasibility of mobile cloud gaming with foveated streaming and the possibilities of extending the work towards Virtual Reality.